\documentclass[aps, prd, twocolumn, showpacs, floatfix,
letterpaper, nofootinbib, superscriptaddress,
longbibliography, 10pt]{revtex4-2}

\usepackage{amsmath, amssymb, amsfonts, physics, bm, ulem}
\usepackage{graphicx}
\usepackage{epsfig}
\usepackage{pgf}
\usepackage{color}
\usepackage[T1]{fontenc}
\usepackage[utf8]{inputenc}
\usepackage[english]{babel}
\usepackage{makecell}
\usepackage{hyperref}
\usepackage{mathtools}
\usepackage{soul}
\usepackage{relsize}
\usepackage{xcolor}

\definecolor{green}{rgb}{0.19,0.64,0.54}
\definecolor{blue}{rgb}{0,0,1}
\definecolor{reddish}{rgb}{0.65, 0.2, 0.2}
\definecolor{darkgreen}{rgb}{0.2,0.7,0.3}
\definecolor{darkblue}{rgb}{0.3,0.40,0.48}
\definecolor{gray}{rgb}{.8,.8,.8}

\hypersetup{,
  pdftitle={TwoFluidBounce},
  pdfauthor={S. Vitenti, P. Peter and N. Pinto-Neto},
  colorlinks=true, 
  linkcolor=darkblue, 
  citecolor=darkgreen, 
  filecolor=reddish, 
  urlcolor=blue, 
  linktocpage=true
}

\newcommand{\ex}{\mathrm{e}}
\newcommand{\GN}{G_\textsc{n}}
\newcommand{\Rez}{\real\mathrm{e}}

\newcommand{\ci}{\mathsf{i}}

\newcommand{\lP}{\ell_\textsc{p}}

\newcommand{\bb}{_\textsc{b}}
\newcommand{\rr}{_\text{r}}
\newcommand{\ww}{_{w}}

\renewcommand{\bar}{\overline}

\newcommand{\gw}{\gamma}
\newcommand{\ec}{\bar{\delta}}

\newcommand{\RH}{R_{H_0}}
\newcommand{\kc}{k_\mathrm{c}}
\newcommand{\rwavelength}{\lambda}

\usepackage{afterpage}
\usepackage{silence}
\usepackage{bbm}
\usepackage{float}
\usepackage{braket}
\usepackage[toc,page]{appendix}
\makeatother
\newcommand{\dpar}[1]{\left(#1 \right)}

\newcommand{\ii}[0]{\mathsf{i}} 

\usepackage{afterpage}
\usepackage{ragged2e}
\newcommand{\secref}[1]{\hyperref[#1]{Section \ref*{#1}. \protect\normalfont\nameref*{#1}}}
\usepackage{comment}

\begin{document}
\title{Two Fluid Quantum Bouncing Cosmology I: Theoretical Model}

\author{Sandro Dias Pinto Vitenti}
\email{vitenti@uel.br}
\affiliation{Departamento de Física, Universidade
  Estadual de Londrina, Rod. Celso Garcia Cid, Km 380, 86057-970,
  Londrina, Paran\'a, Brazil}

\author{Nelson Pinto-Neto}
\email{nelsonpn@cbpf.br}
\affiliation{COSMO -- Centro Brasileiro de Pesquisas
  Físicas, Rua Dr.\ Xavier Sigaud 150,
  22290-180, Rio de Janeiro -- RJ, Brasil}

\author{Patrick Peter}
\email{peter@iap.fr}
\affiliation{$\mathcal{G}\mathbb{R}\varepsilon\mathbb{C}\mathcal{O}$
  -- Institut d'Astrophysique de Paris, CNRS \& Sorbonne
  Université, UMR 7095 98 bis boulevard Arago, 75014 Paris,
  France}

\author{Luiz Felipe Demétrio}
\email{demetrio.luizfelipe.fis@gmail.com}
\affiliation{Departamento de Física, Universidade
  Estadual de Londrina, Rod. Celso Garcia Cid, Km 380, 86057-970,
  Londrina, Paran\'a, Brazil}

\date{\today}

\begin{abstract}
  Bouncing cosmologies offer an alternative to inflation by resolving the initial
  singularity through a contracting phase followed by a bounce into expansion. In many
  such models, the contracting phase is dominated by a single matter component,
  typically pressureless dust, which leads to an almost scale-invariant spectrum of
  scalar cosmological perturbations with a slight blue tilt, so that generating the
  observed red-tilted spectrum within this framework was challenging. In this work, we
  consider a more realistic scenario in which the contracting phase includes both
  matter and radiation, as required on physical grounds. We show that the presence of
  radiation can naturally induce a red tilt in the spectrum of curvature perturbations
  seeded by quantum vacuum fluctuations in the remote past of the contraction. Since
  the perturbations of the two fluids are coupled via gravity, vacuum initial
  conditions must be carefully defined. We demonstrate that, without fine-tuning, the
  resulting entropy perturbations are subdominant with respect to curvature
  perturbations. This suggests that a minimal two-component bounce model, involving
  only ordinary matter and radiation, can connect to the standard expanding cosmology
  with observationally viable initial conditions.
\end{abstract}


\maketitle

\section{Introduction}

The matter bounce scenario~\cite{Brandenberger2002,Wands2004, Peter:2006hx} has been
proposed as a cosmological model in which the universe initiates in a contracting
phase dominated by an almost pressureless fluid, bounces due to quantum effects or
corrections to general relativity (GR), and subsequently connects with the standard
hot big-bang expanding model. A mapping between these models and an almost de Sitter
inflationary expansion have been found~\cite{Wands1999} and explored in
depth~\cite{Falciano2019}. In such a scenario, the spectral index is found to be
$n_\textsc{s}-1=12w/(1+3w)$, where $w=p/\rho$ is the equation of state parameter
relating the fluid pressure $p$ to its energy density $\rho$.

In case the fluid is represented by a canonical scalar field, it is possible to set
$w<0$ in order to obtain a red-tilted spectral index, as well as $|w|\ll 1$ to agree
with the data. Then, the sound speed of the perturbations satisfies $c_\textsc{s}^2=1$
implying, in most cases, that the ratio $r=\mathcal{A}_\textsc{t}
  /\mathcal{A}_\textsc{s}$ between tensor and scalar perturbations amplitudes, is much
larger than that observed. This particular problem can however be addressed as, e.g.,
in Ref.~\cite{Bacalhau:2017hja}, through a relative enhancement of the amplitude of
scalar perturbations due to quantum effects near the bounce. Another feature of this
model is that the scalar field can also act as a late-time dark energy component, but
only during the expanding phase, hence constituting a bouncing model with a dark energy
phase that does not affect the initial conditions posed in the contracting phase.

Assuming the almost pressureless matter is described phenomenologically by a perfect
fluid, sometimes modeled by a $k-$essence scalar field, the sound speed satisfies
$c_\textsc{s}^2=w$, so that tensor perturbations are highly suppressed, satisfying the
observed constraints on $r$. The fluid could also be understood as dark matter, an
observed component of the Universe, which should play an essential role in the
contracting phase of the model. If this is the case, there is no addition of an extra
unobserved scalar field in the cosmological model, as described in the precedent
paragraph and as is required by inflationary models with the inflaton. However, it is
physically implausible that such description can allow $w<0$, leading to obvious
instabilities, and one must conclude the model to be blue-tilted, contrary to
observations.

The aim of this paper is to add some complexity to the matter bounce scenario to make
it more realistic by adding the expected radiation-dominated phase at high
temperatures, and calculate the consequences in the scalar perturbation theory and the
resulting spectral index. We assume both components to be phenomenologically described
by fluids, so our model only contains the matter components already observed, namely
dark (and ordinary) matter and radiation. Note also that, at small scales, the
radiation fluid starts dominating and the background standard model cosmological
scenario naturally emerges after the bounce.

Such a model may a priori be plagued by two issues, namely the spectrum it may produce
and the singularity. Indeed, if a dust fluid naturally lead to a scale invariant
spectrum $n^\text{dust}_\textsc{s}=1$, the part of the spectrum produced during the
radiation-dominated epoch is much worse, leading to the very blue tilt
$n^\text{rad}_\textsc{s}=3$ so that combining both spectra seems very unlikely to
produce a slightly red spectrum $n^\text{eff}_\textsc{s} \lesssim 1$ as measured. We
shall see that in practice, this is exactly what happens. As for the singularity
problem, inclusion of radiation can also help as the emergence of quantum effects
beyond classical GR at small scales can lead replacing it by a regular bounce.

Lacking a satisfactory theory for quantum gravity~\cite{Kiefer:2025udf}, many effective
approaches have been developed to implement a quantum bouncing scenario, like loop
quantum cosmology~\cite{Bojowald:2005epg, Neves:2024pad}, string theory effective
actions~\cite{Brandenberger:2023ver}, and the more straightforward canonical
quantization of gravity~\cite{Chataignier:2023rkq}. In the framework of this last
approach, we have published a series of papers
\cite{Peter:2005hm,Peter:2006id,Peter:2006hx,Falciano:2008nk,Vitenti:2012cx,
  Falciano:2013uaa,Peter:2015zaa} calculating the Hamiltonian constraint up to second
order in perturbation theory over homogeneous and isotropic backgrounds for different
matter fields (including canonical scalar fields and fluids). Using a proper definition
of quantum trajectories~\cite{AcaciodeBarros:1997gy, Pinto-Neto:2013toa,
  Malkiewicz:2019azw} in configuration space, we were able to show that many quantum
trajectory solutions for the background are free of singularities, replaced by a
bounce, and reach the classical limit at large scales. The perturbations obey the same
equations as in the semi-classical theory (in which only the perturbations are
quantized), but with the classical background trajectories substituted by the quantum
ones in the relevant equations.

One must however keep in mind that such an approach is not expected to be reliable very
close to Planckian energies: the minimum curvature scale of the universe evolution,
which is usually reached at the bounce, should thus be sufficiently far from the Planck
length. As the relevant distance scales (in appropriate units) are sufficiently large
even near the bounce, the matter content of the universe can be adequately approximated
by a single perfect fluid with constant equation of state $w_\textsc{b}$ near the
bounce. In practice, in the present case of matter and radiation, one expects radiation
domination during this high temperature dense phase, so that $w_\textsc{b}\to w\rr=1/3$.

In what follows, we first, in Sec.~\ref{Sec:backd}, summarize our model and compute
the quantum corrections to the scale factor evolution as a function of $w\bb$ and the
late time parameters, namely the other fluid equation of state $w$, and the various
relative contributions measured today. Our goal is to describe a model taking care of
the observational knowledge of the cosmological parameters, but extended to account for
earlier contracting and bouncing phases.

On this well-defined phenomenological background, we evaluate, in Sec.~\ref{sec:perts},
the relevant perturbations and their evolution, assuming, in a fashion similar to what
is customarily assumed in inflationary models, an initial quantum vacuum state for this
many-fluid system~\cite{Peter:2015zaa}. It is crucial then to remark that the presence
of two fluids in the matter sector leads to coupled scalar perturbations, for which the
usual techniques for defining vacuum states cannot be applied. To deal with this
problem, we apply the coupled adiabatic vacuum prescription~\cite{Peter:2015zaa} to
define an appropriate vacuum state and then calculate the time evolution of the
relevant observables (power spectra and correlations) to conclude that, even though we
consider a two-fluid situation, the curvature mode largely exceeds the amplitude of the
isocurvature contribution and yields an almost scale-invariant and slightly red
Gaussian spectrum. This last feature stems from the realization that although one
indeed reaches $n^\text{rad}_\textsc{s}=3$ for very small wavelengths, the associated
amplitude is much reduced compared to the scale invariant part coming from the
matter-dominated epoch, so the overall spectrum initially decreases in order to connect
the two components $n^\text{dust}_\textsc{s}=3$ and $n^\text{rad}_\textsc{s}=3$: a
global power-law fit for the full spectrum then naturally yields
$n^\text{fit}_\textsc{s}< 1$, which can then be made to agree with the observed value.

One ends up with a natural extension of the $\Lambda$CDM model: a large, classical,
dust dominated ($w\ll 1$) contracting phase phase smoothly connects by a regular
quantum radiation-dominated bounce to a similar expanding dust era. Such a scenario has
no free parameter except for the value of the scale factor at its minimum, which we
denote by $a\bb \equiv a(\eta\bb)$, with $\eta\bb$ the conformal time at the bounce.
This parameter is subsequently fixed by the microwave background observations as it is
related with the amplitude of the final curvature power spectrum.

\section{Perfect fluid quantum bounce}
\label{Sec:backd}

\subsection{Single Fluid Bounce}
\label{subsec:single_fluid}

As discussed in the introduction, we assume that the relevant distance scales are
sufficiently large during the overall evolution of the universe that one can
coarse-grain its matter content distribution and thus approximate its description by
means of a perfect fluid, keeping in mind that this hypothesis needs be verified when
the relevant parameters are compared with observational data. Assuming natural units
($\hbar=c=1$) throughout, our classical starting point therefore contains the
Einstein-Hilbert GR action $\mathcal{S}_\textsc{he}$ together with that describing a
perfect fluid $\mathcal{S}_\text{fluid}$, namely
\begin{equation}
  \mathcal{S}_\text{tot} = \mathcal{S}_\textsc{he}+\mathcal{S}_\text{fluid} =
  - \int \sqrt{-g} \left( \frac{R}{6\lP^2} + p \right) \dd^4 x,
  \label{action}
\end{equation}
with $\lP=(8\pi\GN/3)^{1/2}$ the Planck length, $\GN$ the gravitational constant, and
$p$ the fluid pressure, related to its energy density $\rho$ through $p=w\bb\rho$; in
the practical application below, we shall set $w\bb\to w\rr=1/3$ to describe a
radiation-dominated bounce. Note at this point that we are considering only one fluid,
as we are interested in describing the quantum-led bounce; other fluids will enter in
the classical regime, and we assume their relative contributions to be negligible
during the quantum phase.

Restricting attention to the flat metric
\begin{equation}
  \dd s^2  = -N^2 (\bar\tau)\dd \bar\tau^2 +
  a^2 (\bar\tau) \delta_{ij}\dd x^{i}\dd x^{j},
  \label{ds2}
\end{equation}
thereby defining a timelike variable $\bar\tau$, the scale factor $a(\bar\tau)$ and the
lapse function $N(\bar\tau)$. Describing the fluid in terms of velocity potentials, the
corresponding Hamiltonian $H_\text{tot}$ resulting from the action \eqref{action} reads
\begin{equation}
  H_\text{tot} \equiv N\left( \frac{P_\tau}{a^{3w\bb}} -
  \frac{\lP^2 P_a^2}{4a V_\text{c}}\right) \approx 0,
  \label{constraint}
\end{equation}
the last weak inequality expressing the classical Dirac constraint. In
Eq.~\eqref{constraint}, $V_\text{c}$ is the (finite) comoving volume of the spatial sections
(space being assumed compact for subsequent quantization reasons) and $P_\tau$ the
canonical momentum conjugate to a variable $\tau$ defined by the fluid (to be later
identified with the time variable $\bar\tau$). The comoving volume $V_\text{c}$ and the Planck
length $\lP$ can be pulled out of the system and absorbed in a variable redefinition.

Setting the lapse $N\to a^{3w\bb}$ in \eqref{constraint} results in a deparametrization
of the Hamiltonian $H_\text{tot}$, whose linear term becomes exactly $P_\tau$. The
associated fluid variable thus takes the meaning of a time so we can identify $\bar\tau
  = \tau$ in what follows. Quantizing {\sl \`a la} Dirac by substituting $P_a \to
  -i\partial/ \partial a$ and $P_\tau \to -i\partial/\partial \tau$ (and hence
$H_\text{tot}\to \hat{H}_\text{tot}$) then transforms the timeless Wheeler DeWitt
equation $\hat{H}_\text{tot} \Psi = 0$ into the time-dependent Schrödinger equation
\begin{equation}
  i \frac{\partial}{\partial \tau}\Psi = \frac{\lP^2}{4V_\text{c}}
  \left\{
  a^{(3w\bb-1)/2}\frac{\partial}{\partial a} \left[
    a^{(3w\bb-1)/2}\frac{\partial}{\partial a}\right]
  \right\}\Psi,
  \label{WDW0}
\end{equation}
for the wave function $\Psi(a,\tau)$, equation in which we chose a specific ordering of
the non commuting operators $a$ and $p_a$ compatible with covariance. Changing the
fundamental variable $a$ to $q$ through
\begin{equation}
  q=\frac{2\sqrt{V_\text{c}}}{3(1-w\bb)\lP} a^{\frac32(1-w\bb)}
  \label{aq}
\end{equation}
transforms Eq.~\eqref{WDW0} into the time-reversed free particle Schrödinger
equation:\footnote{Note that, the scale factor being
  assumed dimensionless, the variable $q$ has dimensions
  given by those of $\sqrt{V_\text{c}}/\lP$, i.e. the square-root of a length.}
\begin{equation}
  i \frac{\partial}{\partial \tau}\Psi(q,\tau) = \frac{1}{4}
  \frac{\partial^2}{\partial q^2} \Psi(q,\tau).
  \label{SchrQ}
\end{equation}
As \eqref{SchrQ} is restricted to positive values of $q$, one needs to impose the
relation
\begin{equation}
  \left. \left( \Psi^{\star}\frac{\partial\Psi}{\partial q}
  -\Psi\frac{\partial\Psi^{\star}}{\partial q}\right)
  \right|_{q=0}=0,
  \label{cond}
\end{equation}
in order to preserve the Hamiltonian's self-adjointness~\cite{Pinto-Neto:2013toa}. This
condition is trivially satisfied by the (real) Gaussian state
\begin{equation}
  \Psi_{\text{ini}} (q) = \left(\frac{8}{\pi\tau\bb}\right)^{1/4}
  \exp\left(-\frac{q^2}{\tau\bb}\right),
  \label{initial}
\end{equation}
which we choose as our initial condition. Using the propagator of Eq.~\eqref{SchrQ}
then yields the full time-dependent wave function $\Psi(q,\tau)$, written in terms of the
rescaled scale factor $\tilde{a}^{\frac32 (1-w\bb)} = \sqrt{V_\text{c}} a^{\frac32 (1-w\bb)}/\lP$, namely
\begin{equation}
  \Psi(\tilde{a},\tau) = \left( \frac{8\tau\bb/\pi}{\tau^2+\tau\bb^2}
  \right)^{1/4} \exp\left[ \frac{-4\tau\bb
      \tilde{a}^{3(1-w\bb)}}{9(1-w\bb)^2\left( \tau^2+\tau\bb^2\right)}\right]
  \ex^{iS},
  \label{solPsi}
\end{equation}
the phase $S(\tilde{a},\tau)$ being given by
\begin{equation}
  S(\tilde{a},\tau) = \frac{\pi}{4} -\frac12 \arctan \left( \frac{\tau\bb}
  {\tau}\right) - \frac{4\tau \tilde{a}^{3(1-w\bb)}}{9(1-w\bb)^2\left(
    \tau^2+\tau\bb^2\right)}.
  \label{phaseS}
\end{equation}
Eq.~\eqref{solPsi} was first obtained in Ref.~\cite{Peter:2006hx} in which it appears
as Eq.~(20), with slightly different notations. Eq.~\eqref{cond}, being proportional to
the conserved current of the Schrödinger equation, is then satisfied at all times.

There are various ways to obtain a meaningful trajectory out of the wave function. Most
are semiclassical (or semiquantum~\cite{Martin:2021dbz}), when the probability density
is sufficiently peaked. In a more general situation however, e.g.~\cite{Mazde:2025zne},
one may use the formulation of Refs.~\cite{deBroglie:1927,Bohm:1951xw,Bohm:1951xx}
adapted to quantum cosmology~\cite{AcaciodeBarros:1998nb}, specifically based on such
trajectories obtained in a quantum equivalent of the eikonal light-ray approximation in
electrodynamics~\cite{Holland:1993ee}. Among the many advantages of such a formulation
is the fact that the "quantumness" of the trajectory is easily evaluated by a simple
computation of the quantum potential $Q[\Psi(\tau)]$ acting on the trajectory which can
be calculated from the wave function itself: the smaller $Q$, the more classical the
system.

In the case at hand, the trajectory is obtained by setting the momenta $P_q = \partial
  S/\partial q$, as in the Hamilton-Jacobi theory, but now $S(q,\tau)$ is the phase of
the wave function. The relationship between the momenta and velocities is obtained as
usual, with $\dot q$ given by the Poisson bracket with the reduced Hamiltonian stemming
from the constraint~\eqref{constraint}, which, in the classical case, would simply be
the partial derivative of this Hamiltonian with respect to $q$, i.e. Hamilton equation.
In the case of quantum cosmology, we get
\begin{equation}
  \frac{\dd \tilde{a}}{\dd \tau} = -\frac{\tilde{a}^{3w\bb-1}}{2}
  \frac{\partial S}{\partial \tilde{a}},
  \label{dadT}
\end{equation}
where $\partial S / \partial \tilde{a}$ is given by
\begin{equation}\label{dSda}
  \frac{\partial S}{\partial \tilde{a}} = -
  \frac{4 \tau \tilde{a}^{2-3w\bb}}{3\left(\tau^2+\tau\bb^2\right)
    \left(1-w\bb\right)}.
\end{equation}
Substituting \eqref{dSda} in \eqref{dadT} then gives the first order ordinary
differential equation
\begin{equation}
  \frac{\dd \tilde{a}}{\dd \tau} = \frac{2}{3}
  \frac{\tau \tilde{a}}{\left(\tau^2+\tau\bb^2\right)
    \left(1-w\bb\right)},
\end{equation}
which is readily integrated. Its solution can be put in the form
\begin{equation}
  a(\tau) = a\bb \left[1+\left(\frac{\tau}{\tau\bb}
    \right)^2\right]^{\frac{1}{3(1-w\bb)}},
  \label{atau}
\end{equation}
where we switched back to $a$ instead of $\tilde{a}$ to recover a dimensionless variable. We note that this solution explicitly replaces the singularity by a bounce at $\tau=0$, with minimal value $a\bb$ of the scale factor.

Note that $\tau$ is related to cosmic time $t$ through $\dd t = a^{3w\bb}\dd \tau$. One
can then calculate the Hubble parameter in terms of the fluid time $\tau$, namely
\begin{equation}
  H(t) \equiv \frac{1}{a}\frac{\dd a}{\dd t} = \frac{2\tau
    a^{-3w\bb}}{3\left(\tau^2+\tau\bb^2\right) \left(1-w\bb\right)},
\end{equation}
which, when expressed in terms of the scale factor, reads
\begin{equation}
  H = \pm \frac{2}{3\left(1-w\bb\right)\tau\bb}
  \frac{1}{a\bb^{3w\bb}}
  \sqrt{\left(\frac{a\bb}{a}\right)^{3(1+w\bb)}
    -\left(\frac{a\bb}{a}\right)^{6}},
\end{equation}
the plus and minus signs respectively describing the expanding and contracting phases.

To simplify the expressions that follow, we introduce the variable $x \equiv a_0/a = 1
  + z$, with $z$ the usual redshift. In terms of $x$, the Hubble parameter can be
written as
\begin{equation}
  H^2 = \frac{4a_0^{-6w\bb} x\bb^{-3(1-w\bb)}}
  {9\left(1-w\bb\right)^2\tau\bb^2}
  \left[{x^{3(1+w\bb)}
        -\frac{x^6}{x\bb^{3(1-w\bb)}}}\right].
  \label{QH2}
\end{equation}
Equation~\eqref{QH2} is nothing but the usual Friedman equation for a perfect fluid
with equation of state $w\bb$, but with an additional term $\propto - x^6$, i.e. the
corrected Friedman equation is phenomenologically equivalent to adding an extra stiff
``quantum matter'' term with negative energy density $\rho_\textsc{q} \propto a^{-6}$.
One can thus rewrite the quantum evolution as
\begin{equation}
  H^2 = \frac{8\pi \GN}{3}\left( \rho - \rho_\textsc{q} \right),
  \label{QH2_2}
\end{equation}
or, setting $\rho = \Omega\bb \rho_\text{crit} x^{3(1+w\bb)}$, with $\Omega\bb$ the
fraction of energy density compared to the critical density $\rho_\text{crit} \equiv 3
  H_0^2 / (8\pi \GN)$, with $H_0$ the Hubble rate today,
\begin{equation}
  \left( \frac{H}{H_0} \right)^2 = \Omega\bb x^{3(1+w\bb)}
  -\Omega_\textsc{q} x^6,
  \label{HH0}
\end{equation}
thereby defining $\Omega_\textsc{q}$. A direct comparison between \eqref{HH0} and
\eqref{QH2} yields
\begin{subequations}
  \label{Omegas}
  \begin{align}
    \Omega\bb = & \frac{4a_0^{-6w\bb}x\bb^{-3(1-w\bb)}}
    {9H_0^2\left(1-w\bb\right)^2\tau\bb^2},
    \label{OW}
    \\ \Omega_\textsc{q} =&
    \frac{4a_0^{-6w\bb}x\bb^{-6(1-w\bb)}}{9H_0^2
    \left(1-w\bb\right)^2\tau\bb^2}
    = \frac{\Omega\bb}{x\bb^{3(1-w\bb)}}.
    \label{OQ}
  \end{align}
\end{subequations}
As one can see from Eq.~\eqref{HH0}, for $w\bb<1$ the quantum part of this equation
becomes negligible for large values of the scale factor $a$, and the classical
evolution is recovered.

Once the fluid density $\Omega\bb$ is determined, Eq.~\eqref{OW} gives a relation
between $x\bb$ and $\tau\bb$, which are the only free parameters introduced by the
quantum trajectory. Therefore, since the fluid density and $H_0$ are potentially
determined by observational data measured long after the bounce occurred, implementing
such a bouncing scenario requires only one new parameter, namely $x\bb$, i.e. the size
of the reference scale factor in units of its minimum bouncing value.

Given the above, one can easily consider extensions to make this quantum bouncing model
closer to the actual universe. Consider some other fluid with equation of state less
than that of the bounce dominating fluid $w\bb$. In this case, given a long enough
contracting phase, the term $\propto x^{3(1+w\bb)}$ in the Friedman equation dominates
over this new fluid throughout the bouncing phase in which the quantum correction may
be relevant. Adding this new fluid is therefore irrelevant as far as the effective
quantum trajectory is concerned, and one can approximate its contribution to be
relevant only during the classical expanding phase.

In modeling the contracting phase, we do not require it to exactly mirror the expanding
one. Processes such as particle creation could alter the composition of the universe
after the bounce, leading to some differences between the two phases. However, we do
not expect a radical change in the overall composition, and some correspondence between
the contracting and expanding branches must be established. To make this connection, we
introduce the reference scale factor $a_0$, defined during the {\sl contracting} phase
as the value for which the comoving Hubble radius equals the present comoving Hubble
radius, i.e. $|H(t_{0-})|^{-1} = |H(t_{0+})|^{-1} = H_0^{-1}$.
For a perfectly symmetric model with the
bounce placed at $t=0$, this would imply $t_{0-} = -t_{0+}$. In a more general,
slightly asymmetric case, these times need not coincide exactly but remain reasonably
close, ensuring a consistent matching between contraction and expansion.

\subsection{Two Fluids with Radiation dominated bounce}

In what follows, and according to the discussion of the previous section, we make the
assumption that the bounce occurs during the radiation dominated phase
($w\bb\to w\rr = 1/3$ in
the previous section), noting that this is the expected behavior in any model
containing matter or dark matter, since these fluids inevitably behave as radiation as
the universe contracts. We also assume that during the classical evolution at large
scales, an almost pressureless fluid, akin to dark matter, becomes dominant. The
sequence of events we have in mind begins with a phase of matter domination (the dust
contraction in the so-called matter bounce scenario) in which the perturbations we will
be interested in are initiated. Then radiation takes over, and the quantum corrections
trigger the end of contraction through a smooth bounce, from which point the standard
cosmological model ensues. A similar background model, with an exactly pressureless
fluid, was solved in Ref.~\cite{Pinto-Neto:2005pwn}.

To describe the bounce with radiation as the dominant component, the Friedmann equation
\eqref{HH0} is specialized by setting $w\rr \to 1/3$ and replacing $\Omega\bb \to
  \Omega_\text{r}$, so that the corresponding term becomes $\Omega_\text{r} x^4$. In this
case, the lapse choice made above simplifies to $N \to a^{3w\rr} = a$, and the fluid
time $\tau$ coincides with the conformal time, $\tau = \eta$.

To ensure consistency with the observed expanding Universe, we incorporate one
additional component, notably a dust contribution modeled as a perfect fluid with a
small equation-of-state parameter ($w \ll 1$). We exclude any dark-energy-like
component (such as a cosmological constant $\Lambda$) for two reasons: (i) its presence
would not qualitatively alter the bounce dynamics (see also
Ref.~\cite{Bacalhau:2017hja} for the case where a dark energy era takes place only at
large scales in the expanding phase), and (ii) it would necessitate a redefinition of
the perturbation vacuum state, as detailed in Refs.~\cite{maier2012bouncing,
  Penna-Lima:2022dmx}. Consequently, our model explicitly assumes the absence of dark
energy during the contracting phase.

In the limit $w \to 0$, the term $\Omega_w x^{3(1+w)}$ reduces to $\Omega_\text{m}
  x^3$.  While radiation dominates for $x \gtrsim \Omega_\text{m}/\Omega_\text{r}$, the
values of $\Omega_\text{m}$ and $\Omega_\text{r}$ can be chosen such that this
transition occurs sufficiently far from the bounce, where the classical approximation
remains valid. Thus, our model is phenomenologically described by the following
modification of Eq.~\eqref{QH2_2}:
\begin{equation*}
  E^2 = \Omega_\text{r} x^4 + \Omega_w x^{3(1+w)} - \Omega_\textsc{q} x^6,
\end{equation*}
where $E^2 \equiv H^2/H_0^2$, and $E \equiv \sqrt{E^2}$ denotes its positive square
root.

The addition of the extra fluid in Eq.~\eqref{OQ} introduces a slight change in our
previous discussion, namely that the value $x_\textsc{b}$ no longer marks the true
bounce point since $E(x_\textsc{b}) \neq 0$. This is however easily resolved by
recalling that $x_\textsc{b}$ is a free parameter depending on the arbitrary reference
scale factor. We can use its normalization freedom to choose $E(x_\textsc{b}) = 0$.
Doing so fixes the quantum contribution as
\begin{equation*}
  \Omega_\textsc{q} = \frac{\Omega_\text{r}}{x_\textsc{b}^2} +
  \frac{\Omega_w}{x_\textsc{b}^{3(1-w)}},
\end{equation*}
yielding a Hubble function of the form
\begin{equation}
  \label{E_f}
  E^2 = \Omega_\text{r} x^4 \qty[1 - \qty(\frac{x}{x_\textsc{b}})^2] +
  \Omega_w x^{3(1+w)} \qty[ 1 - \qty(\frac{x}{x_\textsc{b}})^{3(1-w)}],
\end{equation}
which now obviously satisfies the bounce condition $E(x_\textsc{b}) = 0$. Note that
this reparametrization preserves late-time cosmology, as it reduces to the classical
relation  $E^2 = \Omega_\text{r} x^4+ \Omega_w x^{3(1+w)}$ for $x \ll x_\textsc{b}$.

Using the above parametrization in Eq.~\eqref{OQ}, we get
\begin{equation}
  \frac{x\bb^{-4}}{a_0^2H_0^2\eta\bb^2} =
  \frac{\Omega_\text{r}}{x\bb^2} +
  \frac{\Omega\ww}{x\bb^{3(1-w)}},
  \label{xb}
\end{equation}
where
\begin{equation}
  \eta\bb = \frac{\RH}{a_0\sqrt{\Omega_\text{r} x\bb^2
      + \Omega\ww x\bb^{1+3w}}},
  \label{etab}
\end{equation}
in which we defined the Hubble radius today $\RH \equiv 1/H_0$. Using $\Omega_\text{r}
  = 10^{-7}$, and $\RH \approx 4283 \text{Mpc}$ (with $H_0 = 70 \,\text{km} \cdot
  \text{s}^{-1} \cdot\text{Mpc}^{-1}$), one gets $\eta\bb \approx 4.2\times
  10^{29}\,\text{m}/(a_0x\bb)$. At an arbitrary time, the Hubble radius is
$R_H=\RH/E$.

A key point concerns the range of validity of the Wheeler-DeWitt approach used to
produce the background bounce. Canonical quantization should only be trusted when the
curvature scale at the bounce is well above the Planck length. The curvature radius is
proportional to $a\bb\eta\bb$, which from the above expressions gives
$$
  a\bb\eta\bb \approx 4.2 \left(\frac{10^{30}}{x\bb}\right)^2 \times 10^{-31}\,\text{m}
  \approx 2.6\left(\frac{10^{30}}{x\bb}\right)^2\times 10^4 \, \lP,
$$
Thus, for values around $x\bb\simeq 10^{30}$ the curvature scale at the bounce is many
orders of magnitude larger than $\lP$. This is the regime in which the Wheeler–DeWitt
treatment can be expected to give a reliable semiclassical description of the
background. The same scale also controls the amplitude of the perturbations, so
remaining in the range up to $x\bb\sim10^{30}$ simultaneously satisfies both
requirements.

\begin{figure*}
  \centering
  \includegraphics[width=0.95\linewidth]{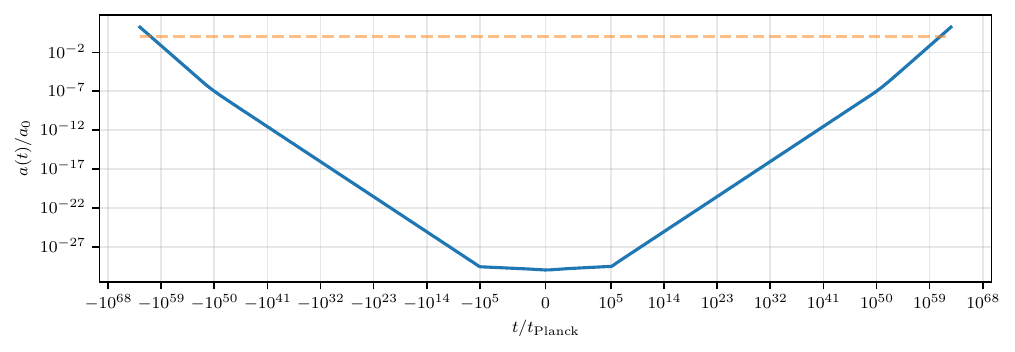}
  \includegraphics[width=0.95\linewidth]{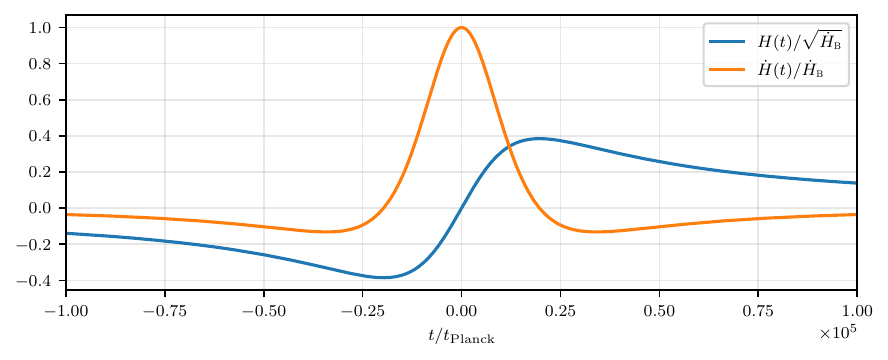}
  \caption{Top: Evolution of the scale factor $a(t)/a_0$ rescaled to its present-day value $a_0$, obtained upon integrating \eqref{E_f} and evaluated as a function of cosmic time $t$ (in units of the Planck time $t_\text{Planck}$ for a symmetric bounce. The contracting phase starts in the matter-dominated era $a\propto \qty(-t)^{2/3}$, transitions to the radiation-dominated era $a\propto \qty(-t)^{1/2}$, and then connects to the quantum bounce (zoomed-in) before going through the same phases in the expanding regime. The dashed orange line represents the size of the Universe today, i.e. $a=a_0$. At its minimum, the scale factor at the bounce is $a\bb =10^{-30}a_0$.\\
  Bottom: Time evolution of the Hubble rate $H(t)$ and its time derivative $\dot{H}(t)$ near the bounce. The blue curve shows the normalized Hubble parameter $H/\sqrt{\dot{H}\bb}$, while the orange curve shows $\dot{H}/\dot{H}\bb$, where $\dot{H}\bb \equiv \dot{H}(0)$ is given by \eqref{dotB}. Cosmic time is measured in Planck units. The Hubble parameter crosses zero at $t=0$, signaling the transition from contraction ($H<0$) to expansion ($H>0$). In a neighborhood of the bounce $\dot{H}>0$, and the behavior of $H(t)$ is approximately linear in $t$, as expected for a smooth, non-degenerate bounce. Far from the bounce both quantities approach the usual power laws of the classical regime. The transition occurs over a time interval of order $10^6$ Planck times, indicating that although the bounce is rapid on cosmological scales, it remains slow compared to the Planck time.}
  \label{fig:placeholder}
\end{figure*}

Unless otherwise stated, throughout this work we adopt the following fiducial
cosmology, i.e. we assume a spatially flat background with
\begin{equation}
  \label{cosmo_bg}
  \begin{split}
    \Omega_\mathrm{r} & = 10^{-7},\quad \Omega_{w} = 1 -
    \Omega_{r},\quad w=10^{-10},                         \\ H_0 & = 70\,
    \text{km}\,\text{s}^{-1}\,\text{Mpc}^{-1}, \quad x\bb = 10^{30}.
  \end{split}
\end{equation}
Figure~\ref{fig:placeholder} shows the explicit solution used in the present work, i.e.
the scale factor $a(t)$, the Hubble rate $H(t)$ and its time derivative $\dot{H}(t)$ as
functions of the cosmic time $t$. These parameter choices, in particular the small
value of $w$ and the large $x_\textsc{b}$ (small $a\bb$), will be discussed in detail
in Sec.~\ref{sec:perts}, where their influence on the perturbations and resulting power
spectra is analyzed.

The linear behavior of the Hubble rate in cosmic time can be shown as follows. Since
Eq.~\eqref{E_f} does not admit a simple analytical solution for the scale factor in
terms of cosmic time, we introduce a convenient time parametrization in which the
Hubble function is directly controlled. Our goal is to show that the transition from
contraction to expansion is smooth and non-degenerate at the bounce. We define a new
lapse function, and hence a new time $\tilde{\tau}$ as
\begin{equation}\label{tau:gauge}
  N \equiv \frac{\dd t}{\dd\tilde{\tau}} = \frac{\tilde{\tau}}{H},
  \qquad \mathrm{sign}(\tilde{\tau})=\mathrm{sign}(H),
\end{equation}
and use the previously defined variable $x$, for which $(\dd x/\dd t)/x = -H$. It then
follows that
\begin{equation}
  \frac{\dd x}{\dd\tilde{\tau}}
  = \frac{\dd x}{\dd t}\frac{\dd t}{\dd\tilde{\tau}}
  = \qty(-Hx)\frac{\tilde{\tau}}{H}
  = -\tilde{\tau} x,
\end{equation}
so that
\begin{equation}
  \frac{1}{x}\frac{\dd x}{\dd\tilde{\tau}} = -\tilde{\tau}
  \quad\Rightarrow\quad
  x(\tilde{\tau})=x_\textsc{b} e^{-\tilde{\tau}^2/2}.
\end{equation}
With this parametrization, Eq.~\eqref{E_f} becomes
\begin{equation}
  \frac{E^2}{\tilde{\tau}^2} =
  \underbrace{\frac{1-e^{-\tilde{\tau}^2}}{\tilde{\tau}^2}}_{\sim 1+
  \mathcal{O}\qty(\tilde{\tau}^2)}\Omega_\mathrm{r} x^4
  + \underbrace{\frac{1-e^{-\frac{3}{2} (1-w)\tilde{\tau}^2}}{\tilde{\tau}^2}}
  _{\sim \frac{3}{2} (1-w)+\mathcal{O}\qty(\tilde{\tau}^2)}\Omega_w x^{3(1+w)}.
  \label{entire}
\end{equation}
For $\Omega_\mathrm{r}>0$, $\Omega_w>0$, and $w<1$, the function \eqref{entire} is
entire in $\tilde{\tau}$ and strictly positive for $\tilde{\tau}\in \mathbb{R}$, so
$E^2/\tilde{\tau}^2$ is itself a strictly positive analytic function. Expanding around
$\tilde{\tau}\approx 0$, one finds $E^2 \approx E_\textsc{b}^2\,\tilde{\tau}^2$, where
\begin{equation}
  E_\textsc{b} \equiv \sqrt{
    \Omega_\text{r} x_\textsc{b}^4 +
    \frac32 (1-w)\Omega_w x_\textsc{b}^{3(1+w)}},
\end{equation}
so that, near $\tilde{\tau}=0$, the Hubble rate behaves linearly with the time
$\tilde{\tau}$, i.e. $H \approx H_0 E_\textsc{b} \tilde{\tau}$, as expected for a
regular non-degenerate bounce. Since $E^2/\tilde{\tau}^2$ is analytic and strictly
positive, $H(\tilde{\tau})$ vanishes only at $\tilde{\tau}=0$, uniquely identifying the
bounce point.

It is easy to see that $\dd H/\dd\tilde{\tau}(\tilde{\tau}=0) = H_0 E\bb > 0$, so the
bounce is indeed a minimum of the scale factor. Since $t$ is monotonic in
$\tilde{\tau}$ near the bounce, because $N=\tilde{\tau}/H$ is finite and nonvanishing
in this limit, this implies $\dd H/\dd t(t\bb)>0$ in cosmic time as well. In fact, even
if the energy densities acquired a smooth, monotonic time dependence (e.g. due to
particle creation), the same conclusion would hold as long as they remain positive and
the equation of state satisfies $w<1$. This property will be useful in the perturbative
analysis below, as it ensures that the equations of motion admit a regular expansion
around the bounce.

In practice, we note that setting the same origin for both times $t$ and
$\tilde{\tau}$, one finds that
\begin{equation}
  H\simeq \qty( H_0 E\bb )^2 t \equiv \dot{H}\bb t,
  \label{dotB}
\end{equation}
as a function of cosmic time. The quantity $\dot{H}\bb$ is used in
Fig.~\ref{fig:placeholder} to rescale $H(t)$ and $\dot{H}(t)$.

\subsection{Normalization to present-day observables}

Having described the parametrization of the contracting phase, we now relate those
parameters to the quantities used to describe the expanding universe and current
observations. Our strategy is to fix the reference scale by matching the magnitude of
the Hubble parameter. Concretely, we define $\tilde a_0$ on the contracting branch and
relate it to the present expanding scale factor $a_0$ through the condition $|H(\tilde
  a_0)| = H(\tilde a_0) \equiv H_0$. This provides a common reference point without
introducing additional arbitrary scales or assuming any dynamical continuity across the
bounce.

In practice, we describe both background solutions by expressing the Hubble parameter
as a function of the scale factor. Since the scale factor alone does not distinguish
between expansion ($H>0$) and contraction ($H<0$), we introduce a notational
convention: quantities without a tilde refer to the contracting branch, while tilded
quantities refer to the expanding one. Accordingly, $H(a)$ denotes the Hubble function
during contraction, and $\tilde H(a)$ the same function evaluated along the expanding
solution.

Both quantities are defined in the standard way as $\dot a/a$, with the tilde serving
only as a bookkeeping device. If one introduces a time coordinate with $t=0$ at the
bounce, this corresponds to $H \equiv (\dot a/a)|_{t<0}$ and $\tilde H \equiv (\dot
  a/a)|_{t>0}$.

During the expanding phase, the background evolution is described by
\begin{equation}
  \frac{\tilde{H}^2}{H_0^2} =
  \tilde{\Omega}_\text{r}\qty(\frac{\tilde{a}_0}{a})^4
  + \tilde{\Omega}_w\qty(\frac{\tilde{a}_0}{a})^3
  + \tilde{\Omega}_\text{b}\qty(\frac{\tilde{a}_0}{a})^3
  + \Omega_\Lambda,
\end{equation}
where quantities with a tilde refer to the expanding branch, and
$\tilde{\Omega}_\text{b}$ and $\tilde{\Omega}_\Lambda$ denote the present-day baryon
density parameter and cosmological constant, respectively. We neglect the small effect
of a nonzero equation-of-state parameter $w$ in the background evolution, i.e. we set
$a^{-3(1+w)} \approx a^{-3}$ since $w\ll 1$.

In contrast, during the contracting phase, we consider models without baryons and
without a cosmological constant. If the only difference between the two branches were
the absence of these components, the corresponding expression would be
\begin{equation}
  \frac{H^2}{H_0^2} =
  \tilde{\Omega}_\text{r}\left(\frac{\tilde{a}_0}{a}\right)^4
  + \tilde{\Omega}_w\left(\frac{\tilde{a}_0}{a}\right)^3.
\end{equation}
Evaluating this expression at $a = a_0$, where by definition $|H| = H_0$, we obtain
\begin{equation}
  1 =
  \tilde{\Omega}_\text{r}\left(\frac{\tilde{a}_0}{a_0}\right)^4
  + \tilde{\Omega}_w\left(\frac{\tilde{a}_0}{a_0}\right)^3.
\end{equation}
This equation corresponds to a fourth-order polynomial for $\tilde{a}_0/a_0$. Rather
than solving it numerically, we exploit the hierarchy $\tilde{\Omega}_\text{r} \ll
  \tilde{\Omega}_w$ and solve perturbatively in powers of $\tilde{\Omega}_\text{r}$. This
yields
\begin{equation}
  \frac{\tilde{a}_0}{a_0}
  = \frac{1}{\tilde{\Omega}_w^{1/3}}
  - \frac{\tilde{\Omega}_\text{r}}{3\tilde{\Omega}_w^{5/3}}
  + \frac{\tilde{\Omega}_\text{r}^2}{3\tilde{\Omega}_w^3}
  - \frac{35\tilde{\Omega}_\text{r}^3}{81\tilde{\Omega}_w^{13/3}}
  + \mathcal{O}\left(\tilde{\Omega}_\text{r}^4\right).
\end{equation}
Using $\tilde{T}_\gamma = 2.7245\,\mathrm{K}$ today and $\tilde{\Omega}_w = 0.25$, we
find $\tilde{a}_0/a_0 \approx 1.6$. Thus, the same value of the Hubble parameter occurs
at a scale factor slightly smaller than today's value in the expanding universe. If the
contracting phase differed from the expanding one only by the absence of baryons and a
cosmological constant, the Hubble function during contraction could be written as
\begin{equation}
  \frac{H^2}{H_0^2} =
  \tilde{\Omega}_\text{r}\left(\frac{\tilde{a}_0}{a_0}\right)^4 x^4
  + \tilde{\Omega}_w\left(\frac{\tilde{a}_0}{a_0}\right)^3 x^3,
\end{equation}
which implies the identifications
\begin{equation}\label{symm}
  \Omega_\text{r} =
  \tilde{\Omega}_\text{r}\left(\frac{\tilde{a}_0}{a_0}\right)^4,
  \qquad \hbox{ and } \qquad
  \Omega_w =
  \tilde{\Omega}_w\left(\frac{\tilde{a}_0}{a_0}\right)^3.
\end{equation}
These identifications hold only under the assumption stated above, namely that the
contracting phase differs from the expanding one solely by the absence of baryons and a
cosmological constant. In this symmetric situation, the radiation content of the
contracting branch is completely fixed by a backward extrapolation of present-day
expanding-phase quantities.

We now relax the symmetric assumption discussed above and allow for more asymmetric
scenarios. In particular, the radiation temperature during contraction need not be fixed
by its expanding-phase value. Moreover, since the density parameters satisfy
$\sum \Omega_i = 1$ independently in each branch, specifying the radiation content during
contraction fully determines the background evolution in that phase. Using the standard
relation $\Omega_\mathrm{r}\propto T_\gamma^4$, evaluated separately in each branch, we
therefore introduce the asymmetry parameter
\begin{equation}
  \gamma_\textsf{assym}
  \equiv
  \frac{T_\gamma}{\tilde{T}_\gamma\left(\tilde{a}_0/a_0\right)}
  =
  \left(\frac{\Omega_\text{r}}{\tilde{\Omega}_\text{r}}\right)^{1/4}
  \left(\frac{a_0}{\tilde{a}_0}\right),
\end{equation}
which quantifies deviations from a symmetric matching between the contracting and
expanding branches. The parameter $\gamma_\textsf{assym}$ measures the mismatch between
the radiation temperature inherited by the contracting branch and the value obtained by
a symmetric extrapolation of today’s CMB temperature back to $a=a_0$, with
$\gamma_\textsf{assym}=1$ corresponding to a perfectly symmetric matching, for which
Eq.~\eqref{symm} is recovered. For reference, the fiducial parameters adopted in
Eq.~\eqref{cosmo_bg} correspond to $\gamma_\textsf{assym} \simeq 0.13$ (using
$\tilde{T}_\gamma$ and $\tilde{\Omega}_w$ as above).

Such an asymmetry is generic in bouncing cosmologies. Entropy production, particle
creation, or incomplete thermalization across the bounce can modify the radiation
content between contraction and expansion, even when the large-scale background
dynamics remain smooth. The parameter $\gamma_\textsf{assym}$ provides a simple and
model-independent way of capturing these effects at the background level, without
committing to a specific microphysical realization of the bounce.

\section{Perturbations in a contracting universe}
\label{sec:perts}

Our primary goal is to connect primordial perturbations to observable signatures in the
cosmic microwave background (CMB). To achieve this, we compute the power spectrum of
comoving modes $\kc$ that seed CMB temperature anisotropies. Recall that we defined
$a_0$ as the scale factor at which the Hubble radius matches its current value; at this
epoch, the corresponding physical wavenumbers are $k = \kc/a_0$. While CMB observations
constrain modes in the range $10^{-6}\,\text{Mpc}^{-1} \lesssim k \lesssim
  1\,\text{Mpc}^{-1}$, we extend our analysis to $10^{-8}\,\text{Mpc}^{-1} \lesssim k
  \lesssim 10^{8}\,\text{Mpc}^{-1}$ to account for potential differences between
contracting and expanding phases and to encompass smaller-scale modes that may become
observable in the future.

In this section, we recall the quantum cosmological perturbations of a two fluid
system~\cite{Peter:2015zaa}. For the fluids part, we follow
Ref.~\cite{Vitenti:2012cx} and apply the variational formalism for perfect fluids
proposed in Ref.~\cite{Schutz:1970my}. Our action is therefore chosen to be
\begin{equation}
  \mathcal{S} = \int \dpar{ \frac{R}{6\lP^2} + p\ww + p\rr }\sqrt{-g}\dd[4] x,
\end{equation}
where $p\ww$ and $p\rr$ respectively refer to the pressures of the almost pressureless
fluid and the radiation fluid leading to the bounce of Sec.~\ref{Sec:backd}. We now
specialize in the case of small perturbations evolving on the flat and regular FLRW
metric introduced in the previous section. We choose comoving coordinates normalized at
$a_0$, the scale factor at which the Hubble radius takes its present value. With this
choice, the determinant of the metric becomes $\sqrt{-g} = (a/a_0)^3$, rather than
$a^3$ as used in~\cite{Peter:2015zaa}.

\subsection{Notations and relevant variables}

The scalar sector is sufficient to capture the physical degrees of freedom relevant to
the dynamics considered here, while vector and tensor modes can be treated
independently. Following the standard decomposition, we write the perturbed metric as
\begin{equation}
  \begin{split}
    \dd s^2 & = -(1-2\phi)N^2\dd\bar\tau^2 - N \partial_i \mathcal{B}\dd x^i \dd\bar\tau                    \\
            & \quad + \left[a^2\delta_{ij}(1+2\psi)-2\partial_i\partial_j\mathcal{E}\right]\dd x^i \dd x^j,
    \label{ds2P}
  \end{split}
\end{equation}
where $\phi$, $\mathcal{B}$, $\psi$, and $\mathcal{E}$ represent the scalar
perturbations of the lapse, shift, and spatial metric, respectively. The perturbations
of the energy–momentum tensor are written similarly
\begin{subequations}
  \begin{align}
    T_0{}^0 & = -\delta\rho,                                              \\
    T_i{}^0 & = N^{-1}(\rho+p)\partial_i \mathcal{V},                     \\
    T_i{}^j & = \delta p \delta_{i}{}^j = c_i^2\delta\rho \delta_{i}{}^j,
  \end{align}
\end{subequations}
where each fluid component carries its own scalar perturbations $(\delta\rho,
  \mathcal{V})$ and speed of sound $c_i$, and we are considering only barotropic fluids
with no internal entropy perturbation.

The variables $\phi$ and $\mathcal{B}$ enter the Lagrangian without time derivatives
and therefore act as Lagrange multipliers enforcing the Hamiltonian and momentum
constraints. Once these constraints are reduced, the remaining degrees of freedom can
be expressed in terms of gauge-invariant combinations of the variables above, which we
define below.

Before analyzing the evolution of perturbations, it is useful to identify and define
the background quantities that most directly enter the perturbation equations. These
quantities also control the validity of adiabatic approximations and determine the
relevant time scales in the system.

In the action, described in detail in Ref.~\cite{Peter:2015zaa}, one can observe that
many terms can be expressed using the combination
\begin{equation}\label{gw}
  \gw_i \equiv \lP^2\frac{a^3 (\rho_i + p_i)}{a_0^3|H|} =
  \frac{(1+w_i)\Omega_i}{\RH}\frac{x^{3w_i}}{E},
\end{equation}
where $i$ denotes the fluid index ($w$ or r).
This combination appears repeatedly in the equations of motion and in the expression of
the canonical variables, particularly in the adiabatic vacuum construction. We
therefore define $\gw_i$ as the {\sl gravitational weight} of fluid $i$, and also
introduce the total gravitational weight $\gw = \gw\ww + \gw\rr$. Rewriting the
perturbation equations in terms of $\gw_i$ and $\gw$ simplifies the formalism and
highlights the contribution of each fluid component to the dynamics.

In addition to the gravitational weights, the sound speeds $c_i$ of each fluid are
relevant background quantities. Since we consider barotropic fluids with constant
equations of state, the individual sound speeds are given by $c\ww = \sqrt{w}$ and
$c\rr = 1/\sqrt{3}$. However, the propagation of perturbations is governed not only by
these values but also by their combinations. The frequencies of the curvature and
isocurvature modes are determined by the effective {\sl curvature} ($c_\zeta$) and {\sl
    isocurvature} ($c_Q$) sound speeds, respectively given by
\begin{align}
  c_\zeta^2 & \equiv \frac{\gw\ww c\ww^2 + \gw\rr c\rr^2}{\gw}, \\
  c_Q^2     & \equiv \frac{\gw\ww c\rr^2 + \gw\rr c\ww^2}{\gw}.
\end{align}

A key dimensionless quantity in the mode equations is the Hubble-scaled wavenumber,
defined as the ratio of the physical wavenumber $(\kc/a)$ to the Hubble rate $|H|$
\begin{equation}\label{Fnu}
  F_\nu \equiv \frac{\kc}{a|H|} = kx R_H,
\end{equation}
where $k \equiv \kc/a_0$ is the physical wavenumber (equivalently, $\rwavelength \equiv
  1/k$ is the reduced wavelength) at the present time. Note that $kx$ is the wavenumber
rescaled to the scale factor $a$, and $R_H$ is the Hubble radius at $a$. Moreover,
although $F_\nu$ does not depend on the fluid type, it does however depend on the scale
factor $x$ in the same way as the gravitational weight $\gw\rr$.

As shown in Ref.~\cite{Peter:2015zaa}, when the Hamiltonian constraints are solved the
system can be described in terms of the curvature perturbation associated with each
fluid, defined by
\begin{equation}
  \zeta_i \equiv \psi + H \mathcal{V}_i,
\end{equation}
where $\mathcal{V}_i$ denotes the (gauge-dependent) velocity potential of fluid $i$ and
$\psi$ the (gauge-dependent) curvature perturbation.\footnote{See Eq.~(33)
  of~\cite{Peter:2015zaa} for the definition of this variable in a flat background. Our
  notation for $\zeta_i$ is consistent with \cite{Vitenti:2012cx, vitenti2012large,
    Peter:2013avv, Peter:2015zaa} and follows Eq.~(5.23) of \cite{Mukhanov1992theory},
  where it is defined as a gauge-invariant combination of metric and fluid perturbations.
  In some of the later literature, this variable is also denoted by $\mathcal{R}$.} One
can define gauge-invariant curvature perturbations for the matter and radiation fluids,
denoted by $\zeta\ww$ and $\zeta\rr$, respectively (see, e.g.,
Ref.~\cite{Peter:2013avv}; they correspond to the fluid velocity perturbations
$\mathcal{U}$ in Ref.~\cite{Peter:2015zaa}). These quantities can be combined into a
total curvature perturbation $\zeta$, defined by the weighted sum
\begin{equation}
  \zeta = \frac{\gw\ww \zeta\ww + \gw\rr \zeta\rr}{\gw},
  \label{zeta}
\end{equation}
the gravitational weights $\gw_i$ being  defined in \eqref{gw}. The conjugate momentum
associated with $\zeta$ involves the gauge-invariant energy density contrasts of each
fluid, given by
\begin{equation}
  \ec_{\rho_i} \equiv \frac{\delta\rho_i}{\rho_i + p_i} + 3\psi,
\end{equation}
where $\psi$ is the metric perturbation defined in Eq.~\eqref{ds2P}. The total energy
density contrast is then expressed as
\begin{equation}\label{ec_rho}
  \ec_\rho = \frac{\gw\ww \ec_{\rho\ww} + \gw\rr \ec_{\rho\rr}}{\gw},
\end{equation}
and the corresponding conjugate momentum for the curvature
perturbation is
\begin{equation}\label{Pizeta}
  \Pi_\zeta = s\gw(3\zeta - \ec_\rho),
\end{equation}
where $s = H / |H|$ denotes the sign of the Hubble parameter, i.e. $s=-1$ in the
contracting phase and $s=+1$ during expansion. This definition of $\Pi_\zeta$ combines
the total energy density fluctuation ($\ec_\rho$) with the spatial curvature ($\zeta$).
Consequently, $\Pi_\zeta$ is dynamically sensitive to curvature perturbations, as
$\zeta$ appears explicitly as the source term in its equation of motion, while pure
isocurvature (entropy) modes do not enter directly. In addition, note that $\Pi_\zeta$
is related to the Bardeen potential $\Psi_k$ in Fourier space through
\begin{equation}\label{def:Psi}
  \Psi_k = -\frac{3xH}{2k^2}\Pi_{\zeta k}.
\end{equation}

In the constant-curvature gauge ($\psi = 0$), the curvature perturbation reduces to $
  \zeta_i = H\mathcal{V}_i$, showing that $\zeta_i$ differs from the velocity potential by
an extra factor of $H$. The physical velocity perturbation is given by the spatial
gradient of $\mathcal{V}_i$. Motivated by this, we introduce the gauge-invariant variable
\begin{equation}
  \label{Vpot}
  F_\nu \zeta_i \quad\xrightarrow{\psi \to 0} \quad k x\, \mathcal{V}_i,
\end{equation}
which, in the constant-curvature gauge, coincides with the modulus of the
gauge-dependent velocity perturbation. Note that in this gauge, the gauge-invariant
contrast $\ec_{\rho_i}$ reduces to the gauge-dependent energy density contrast $\propto
  \delta \rho_i$.

Although the weighted sum \eqref{zeta} gives the total curvature perturbation, the
difference $\zeta\ww-\zeta\rr$ provides the so-called isocurvature mode. Explicitly,
this is~\cite{Peter:2015zaa}
\begin{equation}\label{Smode}
  Q = s\frac{\gw\ww\gw\rr}{\gw} \Delta\zeta, \qquad \Delta\zeta \equiv
  \zeta\rr - \zeta\ww,
\end{equation}
with momentum
\begin{equation}\label{PiS}
  \Pi_Q = \ec_{\rho\ww} - \ec_{\rho\rr},
\end{equation}
which is the usual definition for isocurvature perturbation between two fluids.

In this work, the term ``adiabatic'' is used specifically in its WKB or dynamical
sense, denoting modes whose amplitude evolves slowly compared to their oscillation
period. This is distinct from its other common meaning in cosmology, which describes a
compositional perturbation where the relative number densities of species remain
unperturbed. To avoid conflating these two concepts, we classify perturbations
composition as either curvature (described by $\zeta$ and $\Pi_\zeta$) or isocurvature
(described by $Q$ and $\Pi_Q$). In multi-fluid systems, the curvature perturbation is
sourced not only by the total energy density contrast but also by the collective fluid
velocity potential.  Consequently, isocurvature perturbations can manifest in two
distinct forms: as differences in energy density between fluids (energy isocurvature
$\Pi_Q$) or as differences in their velocity potentials (velocity isocurvature
$\Delta\zeta$).

In terms of the previously defined variables, the Hamiltonian for scalar perturbations
in the two-fluid system, written in Fourier space as $\delta \mathcal{H}^{(2,s)} =
  \sum_k \delta \mathcal{H}^{(2,s)}_k$, is given for each mode $k$ by
\begin{equation}
  \label{2fluidhamiltonian}
  \begin{split}
    \lP^2\frac{\delta \mathcal{H}^{(2,s)}_k}{N\vert H\vert} = & \,
    \frac{c_\zeta^2}{2\gw}\Pi_{\zeta k}^2 + \frac{\gw\ww\gw\rr
    c_Q^2}{2\gw}\Pi_{Q k}^2                                        \\ & +
    s\frac{(c\rr^2-c\ww^2)\gw\rr\gw\ww}{\gw^2}\Pi_{\zeta k}\Pi_{Q k}
    \\ & + \frac{\gw F_\nu^2}{2}\zeta_k^2 + \frac{\gw
      F_\nu^2}{2\gw\rr\gw\ww} Q_k^2,
  \end{split}
\end{equation}
where $N$ is the lapse function, whose choice will be specified below.

Our conventions for the momentum $\Pi_\zeta$ and the variable $Q$ are related to those
in Ref.~\cite{Peter:2015zaa} by the rescaling
$$
  \Pi_\zeta = \lP^2 \Pi^{\text{\cite{Peter:2015zaa}}}_\zeta \qquad Q = \lP^2
  Q^{\text{\cite{Peter:2015zaa}}},
$$
all other perturbation variables being left unchanged. Combined with our definition of
$\gw$, this ensures all quantities are either dimensionless or carry dimensions of
length, even when $\hbar=c=1$ is not imposed, which simplifies the Fourier-space
analysis. This convention introduced an overall factor of $\lP^2$ in the Hamiltonian
above, a factor that can be absorbed through a subsequent rescaling of all perturbation
variables. This rescaling is often adopted to normalize the perturbation variables to
unity and to simplify numerical computations. Whenever required, we restore the factor
by multiplying the perturbation amplitudes by $\lP$.

The two-fluid Hamiltonian in Eq.~\eqref{2fluidhamiltonian} yields the following
Hamilton equations
\begin{subequations}
  \label{TFHamilton}
  \begin{align}
    \label{dotZ}
    \frac{1}{N|H|} \dv{\zeta_k}{\bar{\tau}}       & = \frac{c_\zeta^2}{\gw} \Pi_{\zeta k} + s
    \Delta c^2  \frac{\gw\rr \gw\ww}{\gw^2} \Pi_{Q k},                                        \\
    \label{dotQ}
    \frac{1}{N|H|} \dv{Q_k}{\bar{\tau}}           & = \frac{\gw\ww \gw\rr c_Q^2}{\gw} \Pi_{Q
      k} + s \Delta c^2  \frac{\gw\rr \gw\ww}{\gw^2} \Pi_{\zeta
    k},                                                                                       \\
    \label{dotPZ}
    \frac{1}{N|H|} \dv{\Pi_{\zeta k}}{\bar{\tau}} & = -\gw F_\nu^2 \zeta_k,                   \\
    \label{dotPQ}
    \frac{1}{N|H|} \dv{\Pi_{Q k}}{\bar{\tau}}     & = -\frac{\gw F_\nu^2}{\gw\rr
      \gw\ww} Q_k,
  \end{align}
\end{subequations}
with $\bar{\tau}$ the time coordinate in the perturbed metric \eqref{ds2P}, defined by
the lapse function $N$, and we have set $\Delta c^2 = c\rr^2 - c\ww^2$. The solutions
to Eqs.~\eqref{TFHamilton} will be analyzed below, using vacuum initial conditions
specified in the next section. Before addressing these initial conditions, we derive a
few relevant quantities from the dynamical system that help the physical interpretation
of the perturbation modes, but first we make use of our freedom to choose whatever time
gauge we deem suitable to further simplify the Hamilton equations~\eqref{TFHamilton}.
In particular, we emphasize that Eqs.~\eqref{TFHamilton} contain the combination
$N|H|$, suggesting that a time variable absorbing this factor would be advantageous.

Since we consider an evolution through a full phase of cosmological contraction, with
the scale factor varying over many orders of magnitude, our choice of time coordinate
should reflect this behavior. A natural choice is a logarithmic time variable, which we
denote by $\alpha$, defined through
\begin{equation}
  \label{log_time}
  x = x\bb \exp(-|\alpha|),
\end{equation}
where the bounce occurs at $\alpha = 0$, with $\alpha < 0$ during
contraction and $\alpha > 0$ during expansion. With this choice, the
lapse function becomes
\begin{equation}
  N = \frac{1}{|H|},
\end{equation}
as required. This expression appears ill-defined at the bounce ($H = 0$), so one might
ask whether such a time variable is well defined close to the bounce and if one should
switch to a different time gauge in the vicinity of the bounce. However, a local
analysis of Eq.~\eqref{E_f} shows that $H \propto \pm\sqrt{|\alpha|}$ close to $\alpha
  = 0$, thereby ensuring that the relevant integrals involving $N$ remain
finite;\footnote{Near $\alpha=0$, one has $H \propto \pm\sqrt{|\alpha|}$, so $t = \int
    N\,\mathrm{d}\alpha = \int \mathrm{d}\alpha/H \propto \pm\sqrt{|\alpha|}$; the negative
  branch applies for $\alpha<0$.} the time gauge $\alpha$ is regular. From this point on,
we therefore set $N = 1/|H|$, and denote by a dot a time derivative with respect to the
logarithmic time variable $\alpha$.

\begin{figure}[ht]
  \centering
  \includegraphics[width=0.48\textwidth]{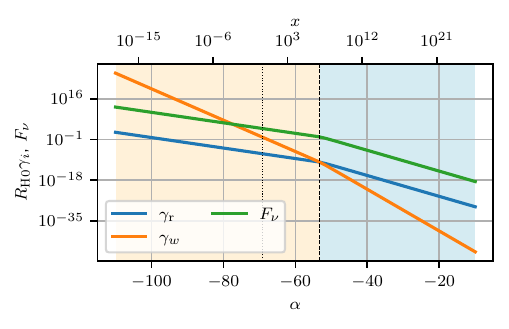}
  \caption{ Evolution of the gravitational weights $R_{H0}\gw\rr$ (blue) and $R_{H0}\gw\ww$
    (orange) as functions of the logarithmic time variable $\alpha$ \eqref{log_time}.
    The top axis shows the corresponding scale factor $x = a_0/a$. The green curve
    plots the Hubble-scaled wavenumber $F_\nu$ \eqref{Fnu} for $k =
      1\,\mathrm{Mpc^{-1}}$, using the fiducial cosmology of Eq.~\eqref{cosmo_bg}. Shaded
    regions mark the radiation-dominated (light blue) and matter-dominated (light
    orange) eras. Vertical lines indicate key transitions: the matter-radiation
    equality (dashed) and the epoch when the Hubble radius matches its present value,
    $x=1$ (dotted). As anticipated from Eq.~\eqref{Fnu}, $F_\nu$ and $\gw\rr$ exhibit
    identical time dependence up to a constant factor. }
  \label{fig:weights}
\end{figure}

The time evolution of perturbations depends on the gravitational weights
$\gw_\mathrm{r}$ and $\gw_w$, which set the relative contributions of radiation and
matter. The ratio $\gw_w/\gw_\mathrm{r}$ tracks the transition from radiation to matter
dominance, while the combination $c_i F_\nu$ determines when each mode enters the WKB
regime as we will discuss in the next section. These quantities remain central in the
adiabatic limit, where derivatives of the background are small, and demonstrate that
the coupling between the two fluids is an intrinsic feature of the system rather than a
negligible factor. In Fig.~\ref{fig:weights}, we show the evolution of
$\gw_\mathrm{r}$, $\gw_w$, and $F_\nu$. All quantities display simple power-law
behavior during the respective domination phases, with a smooth transition around the
matter-radiation equality.

Returning to Hamilton equations~\eqref{TFHamilton}, we identify two main challenges in
solving the system. First, the curvature mode $\zeta_k$ and the entropy mode $Q_k$ are
dynamically coupled. Second, the coefficients in the equations are time-dependent,
reflecting the evolution of the background. As shown in Ref.~\cite{Peter:2015zaa}, at
early times in the contracting phase the system evolves adiabatically: the
characteristic frequencies of the perturbations are much larger than the rates of
change of the background quantities, such as $\gamma_i$ and $F_\nu$.

If the coefficients of the system were constant, an exact diagonalization would reduce
it to independent harmonic oscillators. With time-dependent coefficients, exact
diagonalization is not possible, but we can still perform an {\sl instantaneous}
diagonalization, separating fast and slow degrees of freedom at each time. This
introduces extra terms proportional to time derivatives of the background-dependent
coefficients, which are subdominant in the adiabatic regime. The derivation of the
adiabatic solution is given in Appendix~\ref{app:adiabatic-approximation}.

The analysis reveals two independent degrees of freedom in the system, each associated
with a distinct set of basis functions, respectively called Mode~1, propagating at the
radiation sound speed $c_\mathrm{r}$, and Mode~2, propagating at the matter sound speed
$c_w$. Each set is a nontrivial combination of the two fluid components and cannot be
identified solely with either pure curvature or pure isocurvature perturbations. In the
initial quantum regime, these mode function sets define the corresponding creation and
annihilation operators, ensuring statistical independence. In the classical regime,
their variances correspond to statistical (non-quantum) fluctuations. In the WKB
analysis, each set evolves according to its characteristic sound speed, reflecting the
independent dynamics of the two modes.

To avoid confusion with the underlying fluid components, we adopt the labels Mode~1 and
Mode~2 defined above. This prevents ambiguous notation such as $\zeta_{\mathrm{r}}$,
which could ambiguously refer either to a radiation perturbation or to the total
curvature perturbation in Mode~1's basis functions. In practice, we denote these cases
as $\zeta_{\mathrm{r}1}$ (radiation curvature perturbation in Mode~1) or $\zeta_1$
(total curvature perturbation in Mode~1) respectively.

\subsection{Power Spectra and Correlations}
\label{sec:power_spectra}

The statistical properties of our perturbation modes are fully characterized by two
types of correlation functions: the auto-correlations (power spectra) of individual
fields and their cross-correlations. For the curvature perturbation $\zeta$,
isocurvature perturbation $\Delta\zeta$, and energy density
contrasts, these correlations contain more information about early universe dynamics.

Building on the mode decomposition, we compute both types of correlations under the
vacuum state assumption from Appendix~\ref{app:adiabatic-approximation}. The
statistical independence of Modes~1 and 2 implies that all correlations decompose
additively:
\begin{subequations}
  \begin{align}
    \mathcal{P}_{X}(k)  & = \mathcal{P}_{X1}(k) + \mathcal{P}_{X2}(k),   \\
    \mathcal{P}_{XY}(k) & = \mathcal{P}_{XY1}(k) + \mathcal{P}_{XY2}(k),
  \end{align}
\end{subequations}
where for each mode $i=1,2$ we define:
The \textit{auto-power spectrum} ($X=Y$):
\begin{equation}
  \mathcal{P}_{X i}(k) = \frac{k^3}{2\pi^2} |X_i(k)|^2.
\end{equation}
The \textit{cross-power spectrum} ($X\neq Y$):
\begin{equation}
  \mathcal{P}_{XY i}(k) = \frac{k^3}{2\pi^2} \Rez\left[X_i(k)Y_i^*(k)\right]
\end{equation}
The auto-power spectra (e.g., $\mathcal{P}_{\zeta}$) characterize field covariances,
while the cross-spectra (e.g., $\mathcal{P}_{\zeta\Delta\zeta}$) reveal correlations
between different perturbation types. The $k^3/(2\pi^2)$ normalization ensures these
dimensionless quantities properly represent fluctuation power per logarithmic
$k$-interval, matching CMB analysis conventions.

\begin{figure*}[ht]
  \centering
  \includegraphics{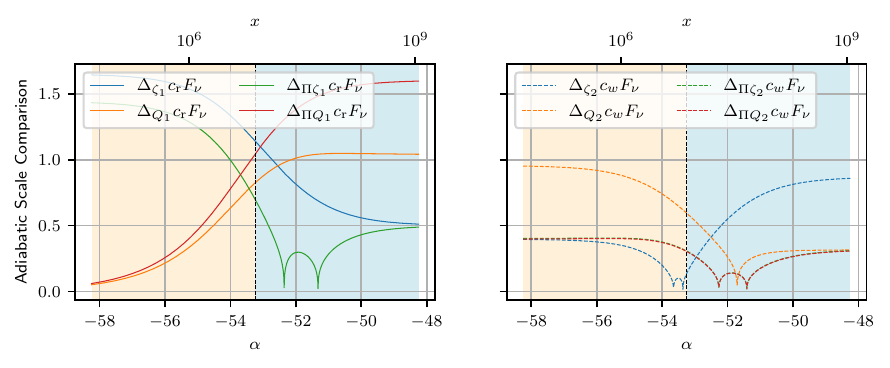}
  \caption{WKB scales $\Delta_{Xi}$ of Eq.~\eqref{eq:adiabatic-scale} for each
    component (Mode~1 left panel, Mode~2 right panel), multiplied by $c_i F_\nu$, as a
    function of the logarithmic time variable $\alpha$ defined by Eq.~\eqref{log_time}
    during the contraction epoch. The upper axis shows the corresponding scale factor
    $x = a_0/a$. Solid lines correspond to Mode~1 and dashed lines to Mode~2. Blue
    denotes $\zeta$, orange denotes $Q$, and green denotes the conjugate momenta
    $\Pi_\zeta$ and $\Pi_Q$. Shaded regions indicate the radiation-dominated (light
    blue) and matter-dominated (light orange) domains.  The vertical dashed line marks
    the matter-radiation transition, Most ratios remain close to unity, except for
    $Q_1$ and $\Pi_{Q1}$, which rise from near zero to unity around the
    matter-radiation transition. The approach to unity indicates the regime where the
    simpler $c_i F_\nu$ correctly approximates the error.}

  \label{fig:adiabatic-scale}
\end{figure*}

\subsection{Sub Sound-Hubble-Scale Regime}
\label{sec:sub-sound-hubble}

For Mode~1, the adiabaticity is controlled by the radiation Sound-Hubble-Scale (SHS)
$c_\mathrm{r} x R_H$, while for Mode~2 it is controlled by the matter SHS $c_w x R_H$.
This naturally divides the time evolution into two separate regimes: super-SHS ($c_i kx
  R_H \ll 1$) and sub-SHS ($c_i kx R_H \gg 1$), each corresponding to the characteristic
SHS of the respective mode.

\begin{figure*}[ht]
  \centering
  \includegraphics{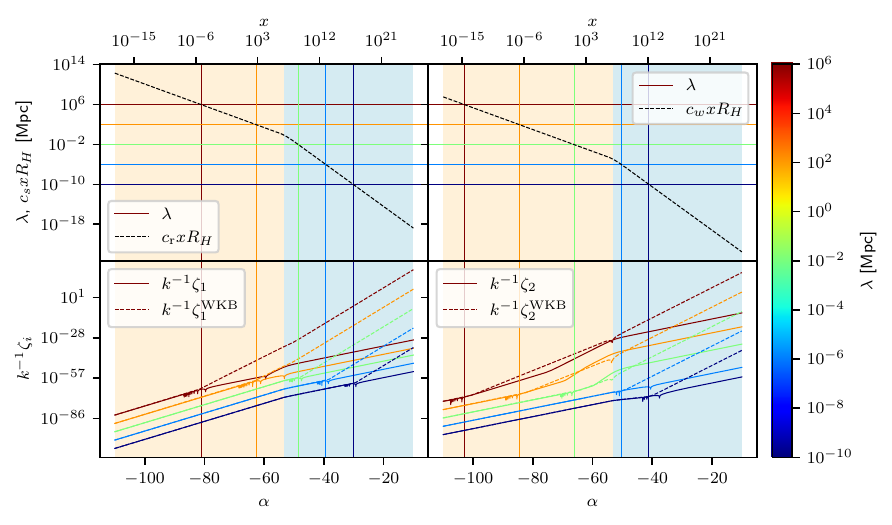}
  \includegraphics{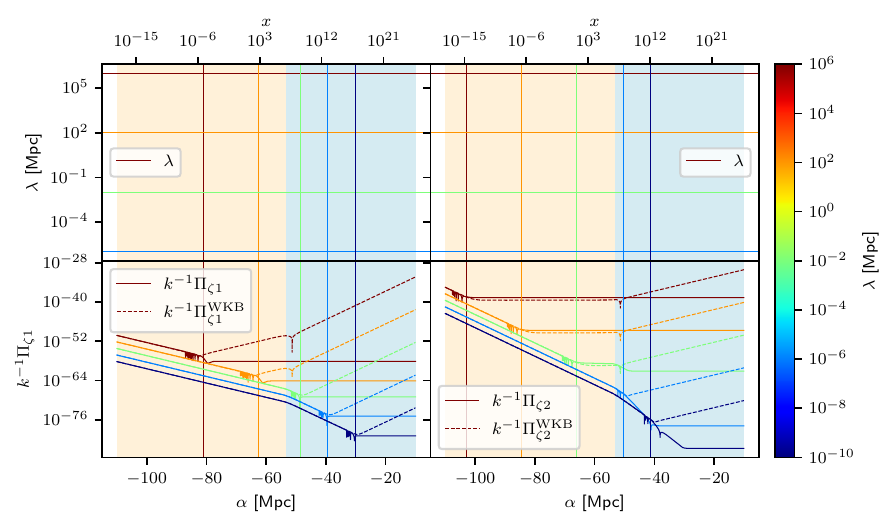}
  \caption{Evolution of multiple modes, each shown in a distinct color with comoving
    wavelength $\rwavelength$ indicated in the color bar (in Mpc). In both figures, the
    top panels show $\rwavelength$ for each mode compared with the comoving Hubble
    radius multiplied by the relevant sound speed: $c_i F_\nu/k = c_i x R_H$ for Mode~1
    (radiation SHS) on the left, and $c_w x R_H$ for Mode~2 (matter SHS) on the right.
    The additional factor $k^{-1}$ separates the modes visually. The bottom panels in
    both figures show the evolution of the corresponding perturbation: $\zeta_1$ and
    $\zeta_2$ in the top figure, and $\Pi_{\zeta 1}$ and $\Pi_{\zeta 2}$ in the bottom
    figure. Vertical lines mark the times when $\rwavelength = c_i x R_H$,
    corresponding to sound-Hubble crossing ($c_i F_\nu = 1$), and help to identify the
    transition between sub- and super-Hubble behavior for each mode. To emphasize
    oscillations inside the Hubble radius (sub-Hubble regime), only the real part of
    the mode functions is plotted. In addition, we also plotted the WKB approximation
    for each mode using the same color but dashed lines. Deep inside this regime,
    computing exact mode functions is computationally expensive, so the adiabatic
    approximation is used instead, which smooths out oscillations and explains their
    absence in that region. The behavior of both $\Pi_{\zeta 1}$ and $\Pi_{\zeta 2}$
    illustrate the approach to the super-SHS solution: Mode~1 reaches the constant
    $\Pi_\zeta$ regime soon after exiting WKB, while Mode~2 shows a delayed approach
    due to its larger effective sound speed ($c_\zeta \simeq 1/3$) relative to $c\ww =
      \sqrt{w} \ll 1$. This contrast clarifies how the crossing of SHS regimes depends on
    the hierarchy between propagation speeds.}
  \label{fig:mode_evolution}
\end{figure*}

In Appendix~\ref{app:adiabatic-approximation}, we derive the first-order adiabatic
corrections to the mode functions, Eqs.~\eqref{eq:delta-1} and \eqref{eq:delta-2}. Our
numerical implementation also includes the second-order corrections, which allow us to
define the WKB scale in Eq.~\eqref{eq:adiabatic-scale} as an estimate of the truncation
error of the adiabatic approximation. While the exact expressions are complicated, they
are well approximated by the simple quantity $1/(c_i F_\nu)$. To illustrate this,
Fig.~\ref{fig:adiabatic-scale} shows the WKB scale for each component multiplied by
$c_i F_\nu$, demonstrating that this approximation is close to unity, varying only
around the matter-radiation transition, but remaining close to unity throughout the
rest of the evolution. The two exceptions are the entropy mode $Q_1$ and its conjugate
momentum $\Pi_{Q1}$, which rise from near zero to unity around the matter-radiation
transition, their smaller values during the matter dominated phase indicate that their
adiabatic corrections are even smaller than those of the other components. Since we
need all components to be in the WKB regime to have a well-defined approximation and
the other components follow $c_i F_\nu$, this indicate the latter is still a good proxy
for the truncation error of the adiabatic approximation.

\begin{figure*}[ht]
  \centering
  \includegraphics{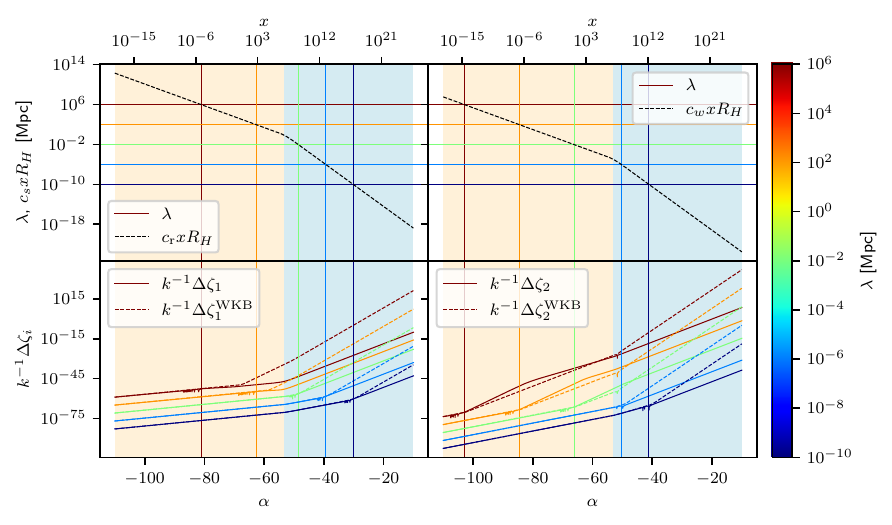}
  \caption{Evolution of the velocity isocurvature perturbation modes $\Delta\zeta_i
      \propto Q_i$ (see Eq.~~\eqref{Smode}), analogous to Fig.~\ref{fig:mode_evolution}
    for the curvature perturbations $\zeta_i$. Top panels: comoving wavelength
    $\rwavelength$ for each mode compared with the comoving Hubble radius multiplied by
    the relevant sound speed, with the extra factor $k^{-1}$ separating the modes
    visually. Bottom panels: evolution of $\Delta\zeta_1$ and $\Delta\zeta_2$ across
    radiation- and matter-dominated eras. Vertical lines indicate sound-Hubble crossing
    ($\rwavelength = c_i x R_H$), and only the real part of the mode functions is
    shown; deep inside this regime the adiabatic approximation smooths out
    oscillations.}
  \label{fig:mode_dzeta_evolution}
\end{figure*}

The system \eqref{TFHamilton} is coupled. This occurs because, even if the fluids are
non-interacting at the background level, their perturbations are coupled through
gravity, leading to the momentum-momentum coupling in
Eq.~\eqref{2fluidhamiltonian}.\footnote{Here only scalars couple to scalars due to the
  standard scalar-vector-tensor decomposition.} Since we assume that cosmological
perturbations originate from quantum vacuum fluctuations in the asymptotic past of the
background model, where the geometry is nearly flat, an appropriate vacuum state must
be defined in this coupled setting. While vacuum state prescriptions (e.g. adiabatic
vacua) are well established for free fields, they are less developed for interacting
fields.  One must therefore employ more general quantization techniques that reduce to
the usual adiabatic construction in the free-field limit, but remain valid for systems
like \eqref{TFHamilton}.

\begin{figure*}[ht]
  \centering
  \includegraphics{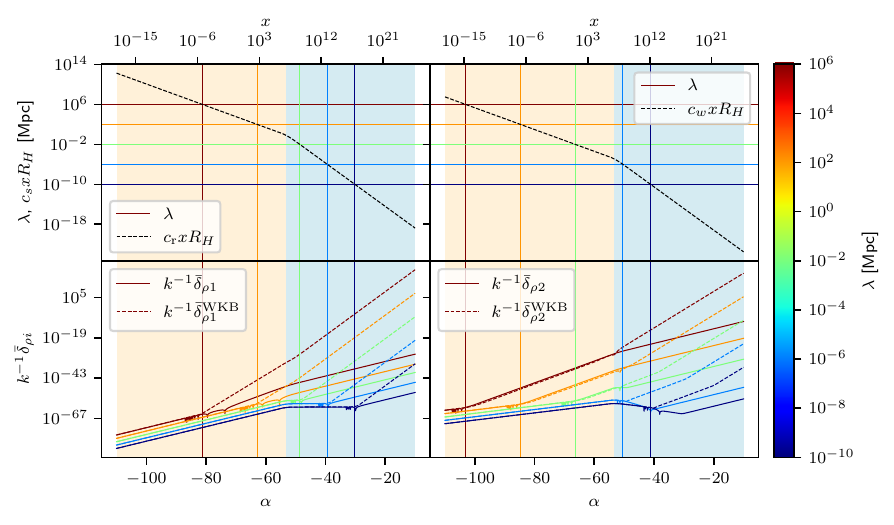}
  \caption{Evolution of the total energy density contrast $\ec_{\rho}$, analogous to
    the previous figures. Top panels: comoving wavelength $\rwavelength$ for each mode
    compared with the comoving Hubble radius multiplied by the relevant sound speed,
    with $k^{-1}$ separating the modes visually. Bottom panels: evolution of
    $\ec_{\rho,1}$ and $\ec_{\rho,2}$ across radiation- and matter-dominated eras.
    Vertical lines indicate sound-Hubble crossing ($\rwavelength = c_i x R_H$), and
    only the real part of the mode functions is shown; deep inside this regime the
    adiabatic approximation smooths out oscillations.}
  \label{fig:mode_drho_evolution}
\end{figure*}

To address this, in Appendix~\ref{app:adiabatic-approximation} we construct initial
conditions by defining a vacuum state for perturbations in the contracting phase, along
the lines of Ref.~\cite{Peter:2015zaa}. There, the generalization of the adiabatic
vacuum to the case of $N$ interacting fluids was first proposed. For clarity, in the
present work we rederive the construction explicitly for the two-fluid system,
diagonalizing the Hamiltonian \eqref{2fluidhamiltonian} in the adiabatic limit and then
building the corresponding adiabatic vacuum in the diagonal basis. This procedure
yields a consistent choice of initial state that extends the standard adiabatic vacuum
to coupled systems, and, to the best of our knowledge, provides a unique treatment in
the literature.

\begin{figure*}
  \centering
  \includegraphics{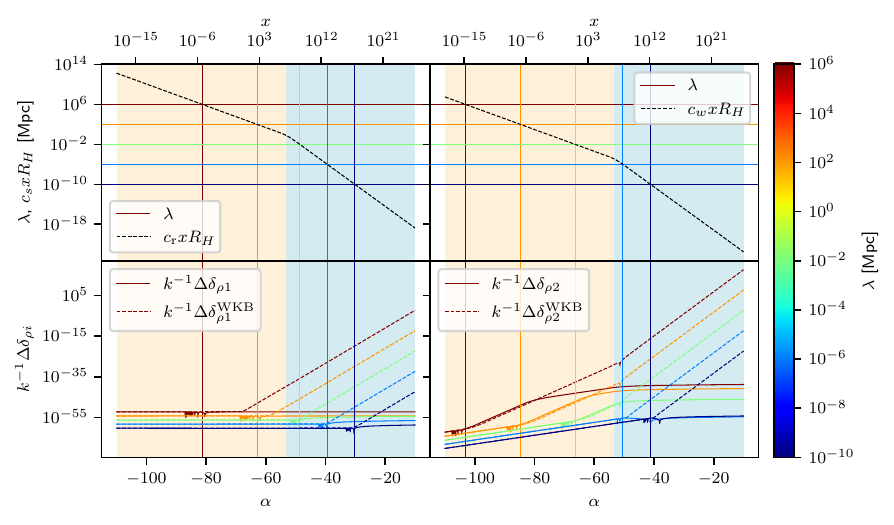}
  \caption{Evolution of the difference between the individual fluid energy density
    contrasts, $\Pi_{Q_i} = \ec_{\rho_\mathrm{r}i} - \ec_{\rho_w i}$, analogous to
    Fig.~\ref{fig:mode_drho_evolution} for the total energy density contrast. Top
    panels: comoving wavelength $\rwavelength$ compared with the comoving Hubble radius
    times the relevant sound speed, with $k^{-1}$ separating the modes visually. Bottom
    panels: evolution of $\ec_{\rho_\mathrm{r},i} - \ec_{\rho_w,i}$ across radiation-
    and matter-dominated eras. Vertical lines indicate sound-Hubble crossing
    ($\rwavelength = c_i x R_H$), and only the real part of the mode functions is
    shown; deep inside this regime the adiabatic approximation smooths out
    oscillations. }
  \label{fig:mode_ddelta_rho_evolution}
\end{figure*}

\subsection{Super Sound-Hubble-Scale Evolution}
\label{sec:super_sound_hubble}

In the super-SHS regime, we expand the Hamilton equations \eqref{TFHamilton}
in powers of $k$. From Eqs.~\eqref{dotPZ} and \eqref{dotPQ}, we see that the
leading-order contributions correspond to constant $\Pi_\zeta$ and $\Pi_Q$.
Consequently, the solutions for $\zeta$ and $Q$ are obtained by integrating these
coefficients. These integrals with $n$ indices $i_1, \cdots i_n$, can be written
compactly as
\begin{equation}
  \mathcal{I}_{i_1\cdots i_n} = \int_{\alpha_1}^\alpha \dd{\alpha'}
  \frac{\gw_{i_1}(\alpha')\cdots \gw_{i_n}(\alpha')}{\gw^2(\alpha')},
\end{equation}
where $\alpha_1$ is an arbitrary initial time. Using these integrals, the leading-order
evolution of $\zeta$ and $Q$ in the super sound-Hubble regime can be expressed as
\begin{subequations}
  \label{super_sound_hubble}
  \begin{align}
    \begin{split}
      \zeta(\alpha) = & \ \zeta(\alpha_1) \\ & + c^2\rr
      \qty[\Pi_\zeta(\alpha_1)\mathcal{I}\rr+\Pi_Q(\alpha_1)s
      \mathcal{I}_{\mathrm{r}w}]          \\ & +c^2\ww
      \qty[\Pi_\zeta(\alpha_1)\mathcal{I}\ww-
      \Pi_Q(\alpha_1)s\mathcal{I}_{\mathrm{r}w}],
    \end{split} \\
    \begin{split}
      Q(\alpha) = & \ Q(\alpha_1)                    \\ & + c^2\rr
      \qty[\Pi_Q(\alpha_1)\mathcal{I}_{\mathrm{r}ww}+
      \Pi_\zeta(\alpha_1)s\mathcal{I}_{\mathrm{r}w}] \\ &
      +c^2\ww
      \qty[\Pi_Q(\alpha_1)\mathcal{I}_{\mathrm{r}\mathrm{r}w}-
      \Pi_\zeta(\alpha_1)s\mathcal{I}_{\mathrm{r}w}],
    \end{split}
  \end{align}
\end{subequations}
These expressions make explicit how the initial momenta, $\Pi_\zeta(\alpha_1)$ and
$\Pi_Q(\alpha_1)$, control the leading-order growth of the perturbations. Each term
corresponds to contributions from the radiation and matter components, weighted by
their respective sound speeds and integrals of the Hamiltonian coefficients.

Now, since the solution derived above is valid only for $F_\nu < 1$, we match it to the
WKB solution at the transition point $c_i F_\nu = 1$, with $i = \mathrm{r}$ for Mode~1
and $i = w$ for Mode~2. This matching slightly extrapolates the approximate solutions
but is sufficient to capture the adiabatic behavior of the perturbations and to connect
smoothly with the super sound-Hubble regime. The end condition for the sub sound-Hubble
solution is therefore
\begin{equation}
  c_iF_\nu = 1, \quad\to\quad k = \frac{1}{c_i x^\times_i(k)
    R_H\qty[x^\times_i(k)]},
\end{equation}
which implicitly defines the end time $x^\times_i(k)$ for each wavenumber $k$.

If this condition occurs deep in the matter-dominated phase, the end time can be
approximated by
\begin{equation}
  x^\times_i(k) \approx \left(\frac{c_i \RH
    k}{\sqrt{\Omega\ww}}\right)^{2/(1+3w)}.
\end{equation}
At the same epoch the background functions reduce to
\begin{align}
  F_\nu  & \approx \frac{\RH k}{\sqrt{\Omega\ww}}\frac{1}{x^{(1+3w)/2}}, \\
  \gw\ww & \approx
  \frac{(1+w)\sqrt{\Omega\ww}}{\RH}\frac{1}{x^{3(1-w)/2}},               \\
  \gw\rr & \approx
  \frac{4\Omega\rr}{3\RH\sqrt{\Omega\ww}}\frac{1}{x^{(1+3w)/2}}.
\end{align}
Inserting these expressions into the leading-order WKB solutions
\eqref{ev1}--\eqref{ev2}, we find
\begin{align}
  \bar{\Pi}_{\zeta 1} & = \ii s\sqrt{\frac{\gw\rr F_\nu}{2c\rr}}
  \approx \ii s\sqrt{\frac{2\Omega\rr}{3\Omega\ww c\rr}\frac{k}{x^{(1+3w)}}}, \\
  \bar{\Pi}_{\zeta 2} & =
  -\ci s\sqrt{\frac{\gw\ww F_\nu}{2c\ww}} \approx
  -\ci s\sqrt{\frac{(1+w)}{2c\ww}\frac{k}{x^2}}.
\end{align}

Evaluating these at the end of the sub-SHS regime gives
\begin{subequations}
  \label{subshs}
  \begin{align}
    \bar{\Pi}^\times_{\zeta 1} & \approx \ci s\sqrt{\frac{2\Omega\rr}{3
        c\rr^3\RH}\frac{(\RH k)^2}{(\RH k)^3}},
    \\ \bar{\Pi}^\times_{\zeta 2} & \approx
    -\ci s\sqrt{\frac{(1+w)}{2c\ww\RH}
    \left(\frac{\Omega\ww}{c\ww^2}\right)^{\frac{2}{1+3w}}\frac{(\RH
      k)^{\frac{12w}{1+3w}}}{(\RH k)^3}},
  \end{align}
\end{subequations}
where superscript $^\times$ indicates the value at the end of the appropriate sub-SHS
crossing time, e.g., $\bar{\Pi}^\times_{\zeta 1} \equiv \bar{\Pi}_{\zeta
    1}\qty[x^\times\rr(k)]$ and $\bar{\Pi}^\times_{\zeta 2} \equiv \bar{\Pi}_{\zeta
    2}\qty[x^\times\ww(k)]$. The corresponding power spectra scale as
\begin{align}\label{ps-mode-1}
  k^3\left\vert\bar{\Pi}^\times_{\zeta 1}\right\vert^2 & \propto k^{2}, \\
  k^3\left\vert\bar{\Pi}^\times_{\zeta 2}\right\vert^2 &
  \propto k^{\frac{12w}{1+3w}}.
\end{align}
Hence, only Mode~2 produces an almost scale-invariant spectrum, while Mode~1 has a
steep blue spectrum with effective spectral index $n = 3$. Which mode dominates depends
on the coefficients in Eq.~\eqref{subshs}. Mode~2 has a larger prefactor due to the
presence of $1/c\ww^5$ and the presence of $\Omega\rr$ on Mode~1's prefactor. On the
other hand, the spectrum of Mode~1 grows with $k^2$ while Mode~2 grows with
$k^{\frac{12w}{1+3w}}$ which is a smaller exponent than $2$ for $w \ll 1$. Thus, up to
$k=1/\RH$, Mode~2 dominates and then Mode~1 starts growing faster and would eventually
overtake Mode~2. However, in practice, this overtaking occurs for modes that exit the
WKB regime during or after the epoch of matter domination, and thus the approximations
above will no longer hold. Nonetheless, for $k\leq 1/\RH$, Mode~2 provides a good
approximation to the spectrum and we will use it to anchor the power spectrum at
$k=1/\RH$.

In our bounce scenario, contraction proceeds up to $x_\textsc{b} = 10^{30}$ well inside
the radiation-dominated era. Since $\gamma_i$ is always a decreasing function of
$\alpha$, the integrands of $\mathcal{I}_i$ increase monotonically with time. In any
single fluid dominated phases, they reduce to simple powers of $x$, so the integrals
can be split as
\begin{equation}
  \int_{\alpha^\times_i(k)}^{0}\frac{\gw_i}{\gw^2}\dd\alpha =
  \int_{\alpha^\times_i(k)}^{\alpha^*} \frac{\gw_i}{\gw^2}\dd\alpha +
  \int_{\alpha^*}^{0}\frac{\gw_i}{\gw^2}\dd\alpha,
\end{equation}
where $\alpha^\times_i(k) = -s\ln\qty[x^\times_i(k)/x_\textsc{b}]$. The first term is bounded above by
\begin{equation}
  \begin{split}
    \int_{\alpha^\times_i(k)}^{\alpha^*}\frac{\gw_i}{\gw^2}\dd\alpha &
    \leq \left[\alpha^*-\alpha^\times_i(k)\right]
    \frac{\gw_i(\alpha^*)}{\gw^2(\alpha^*)}                            \\ & \lesssim
    \left[\alpha^*-\alpha^\times_i(k)\right]\frac{3\RH}{4\sqrt{\Omega\rr}}x^*.
  \end{split}
\end{equation}
Here, $\alpha^*$ is chosen such that the system is deep within the radiation-dominated
phase. In this regime, the background functions simplify to
\begin{align}
  F_\nu & \approx \frac{\RH k}{\sqrt{\Omega\rr}}\frac{1}{x}, \\ \gw\ww
        & \approx
  \frac{(1+w)\Omega\ww}{\RH\sqrt{\Omega\rr}}\frac{1}{x^{2-3w}},
  \\ \gw\rr & \approx \frac{4\sqrt{\Omega\rr}}{3\RH}\frac{1}{x}.
\end{align}
Accordingly, the late-time integrals are
\begin{align}
  \mathcal{I}\rr \approx
  \int_{\alpha^*}^{0}\frac{\gw\rr}{\gw^2}\dd{\alpha} & \approx
  \frac{3\RH}{4\sqrt{\Omega\rr}}(x_\textsc{b} - x^*),          \\ \mathcal{I}\ww \approx
  \int_{\alpha^*}^{0}\frac{\gw\ww}{\gw^2}\dd\alpha   & \approx
  \frac{9\RH(1+w)\Omega\ww}{16\Omega\rr^{\frac32}}\left(\frac{x^{3w}_\textsc{b}
    - x^{*3w}}{3w}\right).
\end{align}
Choosing, for instance, $x^* = 10^{10}\ll x_\textsc{b} = 10^{30}$ (corresponding to the
nucleosynthesis epoch)  has two key effects: it ensures the first integral piece is
negligible compared to the second, and it guarantees the background is in a pure
radiation-dominated phase, allowing the second integral to be evaluated using the
simple power-law approximations. Consequently, we find the late-time integrals are
dominated by the endpoint at $x_\textsc{b}$, exhibiting negligible sensitivity to the
transition time $\alpha^\times_i(k)$. This conclusion holds for all wavelengths that
exit the sound-Hubble radius significantly before the end of contraction.

The integrands of all other integrals decrease monotonically in the radiation-dominated
era, rendering them insensitive to the physics at the bounce scale. Consequently, the
estimates derived in this section are all that is required to compute the super-SHS
approximations~\eqref{super_sound_hubble}. In practice, we estimate the large-scale
power spectrum of $\zeta$ using
\begin{equation}
  \zeta(\alpha=0) \approx c\rr^2 \mathcal{I}\rr \Pi_{\zeta 2}^\times,
\end{equation}
leading to
\begin{equation}
  \label{Pk-zeta-approx}
  \begin{split}
    \mathcal{P}_{\zeta} & = \frac{\lP^2k^3}{2\pi^2}c\rr^4
    \mathcal{I}\rr^2 \vert\Pi_{\zeta 2}^\times\vert^2,    \\ & =
    \frac{\lP^2}{\RH^2}\frac{(1+w)x_\textsc{b}^2\Omega\ww^{2/(1+3w)}}{64\pi^2\Omega\rr
    c\ww^{(5+3w)/(1+3w)}}(\RH k)^{\frac{12w}{1+3w}},      \\ & = 2\times
    10^{-32}\left(\frac{10^{-5}}{c\ww}\right)^{\frac{5+3w}{1+3w}} (\RH
    k)^{\frac{12w}{1+3w}}
  \end{split}
\end{equation}
where we reintroduced the $\lP^2$ factor from \eqref{TFHamilton}. In the last equality,
we used the fiducial cosmology parameters from~\eqref{cosmo_bg}, but we treat the
factor $c\ww$ as a free parameter. Note that the amplitude is very small compared to
the value expected from CMB observations. To achieve a power spectrum amplitude
comparable to CMB observations, the value of $c\ww$ would need to be decreased to
approximately $10^{-10}$. We do not use such values in our numerical integration
because it becomes unstable for $c\ww \lesssim 10^{-5}$. However, for such small values
of $c\ww$, all relevant quantities can be extrapolated from the numerics.

The exceedingly small amplitude of $\mathcal{P}_{\zeta}$ stems from the prefactor
$(\lP/\RH)^2 \propto 10^{-121}$, which arises from imposing quantum initial conditions
on cosmological scales. While the shear and expansion-rate perturbations possess a
runaway growing mode during contraction, increasing as the scale factor decreases $a
  \to a_{\textsc b}$ (equivalently $x \to x_{\textsc b}$) and thereby enhancing the
amplitude, we are ultimately constrained by the requirement that the background
curvature scale remain well above the Planck length $\lP$. This limits the maximum
value of $x_\textsc{b}$. It is important to note that a comoving wavelength $\lambda =
  \RH$ (a cosmological scale today) corresponds to a physical length of approximately
$10^{-4}$m at the bounce, which is still significantly larger than $\lP$. Thus, the
limitation on $x_\textsc{b}$ is not imposed by the perturbations becoming
trans-Planckian, but by the risk of the background curvature itself entering the full
quantum gravity regime which would render inappropriate our use of the Wheeler DeWitt
equation to describe the bounce.

It is important to stress that the super-SHS approximation can only be trusted when the
relevant integral is dominated by one of its endpoints, i.e. for the initial time or at
the bounce ($\alpha=0$). In general, we encounter two types of integrands,
monotonically increasing or decreasing. In both cases the integrand evolves across many
orders of magnitude within the time interval considered, but with different behaviors.
For monotonically decreasing integrands, the dominant contribution comes from the
initial time, where the super-SHS approximation is not very accurate. This is because
at the starting point, the evolution matrix~\eqref{evolmatrix} has Frobenius norm
$\lvert M^n\rvert > 1$ for all $n$, so the approximation is only slowly converging to
the exact solution. The norm falls below unity only later in the evolution, which
explains why the approximation performs better for monotonically increasing integrals.
In this case, not only does the endpoint at $x_\textsc{b}$ provide the dominant
contribution, but also, at the bounce time, one finds $M(\alpha=0)^n \ll 1$ for $n>1$,
ensuring that the initial-time contributions are strongly suppressed. Even so, accuracy
is not guaranteed, since residual early-time evolution can significantly affect the
constant part of the result. A concrete example is provided by curvature perturbations
that cross the SHS close to or after equality, where the large value $F_\nu = 1/c\ww
  \gg 1$ for Mode~2 makes the sensitivity to the early-time dynamics manifest.

The super-SHS dynamics are especially relevant for Mode~2. This stems from the fact
that the sound speed of the matter fluid is very small, $c\ww \ll 1$. Consequently, the
WKB regime for this mode ends when $F_\nu = 1/c\ww \gg 1$. Moreover, Mode~2 is the
dominant contributor to the large-scale power spectrum, so its evolution is of
particular interest.

The super-SHS approximation follows from rewriting $\dot y = M y$ as the integral
equation $y(t) = y(t_0) + \int_{t_0}^t M(t')\, y(t')\, dt'$ and iterating it, so the
expansion involves powers of the evolution matrix $M$. These powers contain terms of
the form (cf. Eq.~\eqref{TFHamilton})
\begin{equation}
  F_\nu^2 c_\zeta^2 \propto \frac{c_\zeta^2}{c\ww^2}.
\end{equation}
In this regime one has $c_i F_\nu = \varepsilon \ll 1$, and the approximation assumes
that successive powers of $M$ remain small. However, the super-SHS series includes
the factor $F_\nu^2 c_\zeta^2 = \varepsilon\, c_\zeta^2/c_i^2$. For the first terms to
be reliable one therefore needs
$$
  \varepsilon\,\frac{c_\zeta^2}{c_i^2} \lesssim 1.
$$

Early in the super-SHS regime, $\varepsilon$ is only slightly below unity. If the ratio
$c_\zeta^2/c_i^2$ is large at that time, the approximation using the first terms breaks
down; it only becomes accurate later, when $\varepsilon$ decreases.

This behavior is visible in Fig.~\ref{fig:mode_evolution}. For Mode~1 the ratio
$c_\zeta^2 / c_r^2$ remains controlled: it begins small ($c_\zeta \simeq c_w$) and
later approaches $c_r$, staying of order unity. As a result, $\Pi_{\zeta 1}$ quickly
approaches its constant super-SHS value after the WKB regime ends. For Mode~2,
$c_\zeta^2 / c_w^2$ is of order unity only at the very beginning of the super-SHS
regime but grows rapidly as $c_\zeta$ approaches $c_r$. Consequently, $\Pi_{\zeta 2}$
continues to evolve for some time after leaving the WKB regime and only later settles
to a constant value. Notably, this early super-SHS evolution has a direct impact on the
final value of $\Pi_{\zeta 2}$ and is responsible for the corresponding reduction in
the power-spectrum amplitude of $\zeta_2$.

Thus, when the total sound speed is still governed by the matter component ($c_\zeta
  \approx c_w$) at WKB exit, the relevant contribution remains of order unity and the
approximation yields a reasonable estimate for the power-spectrum amplitude. At later
times, when radiation determines the sound speed ($c_\zeta \approx c_r$), the same
factor scales as $c_r^2/c_w^2 \gg 1$, leading to a situation where the first terms
would not provide a good approximation and would require many terms to describe this
time interval. Note however that numerically this regime is very well behaved and we
compute it directly.

Therefore, we expect a reliable prediction only for modes that exit the WKB regime
before the matter-radiation sound speed equality, defined by $c\rr^2\gw\rr^2 =
  c\ww^2\gw\ww^2$. This condition corresponds to the scale
\begin{equation}
  x_\textsc{s} = \qty[\frac{c\ww^2}{c\rr^2}
    \frac{3(1+w)\Omega\ww}{4\Omega\rr}]^{1/(1-3w)}.
\end{equation}
Using the definition of $F_\nu$, we obtain the associated wavenumber
\begin{equation}\label{defkS2}
  k^\textsc{s}_2 \equiv \frac{E(x_\textsc{s})}{x_\textsc{s} c\ww \RH}
  = \frac{\sqrt{3}\Omega\ww}{2c\rr\sqrt{\Omega\rr}\RH} +
  \mathcal{O}(w).
\end{equation}

It is important to note that $x_\textsc{s}$ lies within the matter-dominated era. Thus,
while the spectrum is affected by super-SHS evolution, the WKB regime still ends during
matter domination. A second relevant scale for Mode~2 is $\gamma_r=\gamma_w$, given by
\begin{equation}
  x_\mathrm{eq} =
  \qty[\frac{3(1+w)\Omega\ww}{4\Omega\rr}]^{1/(1-3w)}.
\end{equation}
The corresponding wavenumber, obtained again from Eq.~\eqref{Fnu}, is
\begin{equation}\label{defkeq2}
  k^\mathrm{eq}_2 \equiv \frac{E(x_\mathrm{eq})}{x_\mathrm{eq} c\ww
    \RH} =
  \frac{1}{c\ww}\frac{\sqrt{21}\Omega\ww}{4\sqrt{\Omega\rr}\RH} +
  \mathcal{O}(w).
\end{equation}

Beyond this scale, we no longer expect the nearly scale-invariant spectrum derived
earlier to hold. The scales identified above, together with the predicted behavior of
Mode~2 in the super-SHS regime, will be directly tested in the numerical analysis that
follows.

\begin{figure*}
  \centering
  \includegraphics{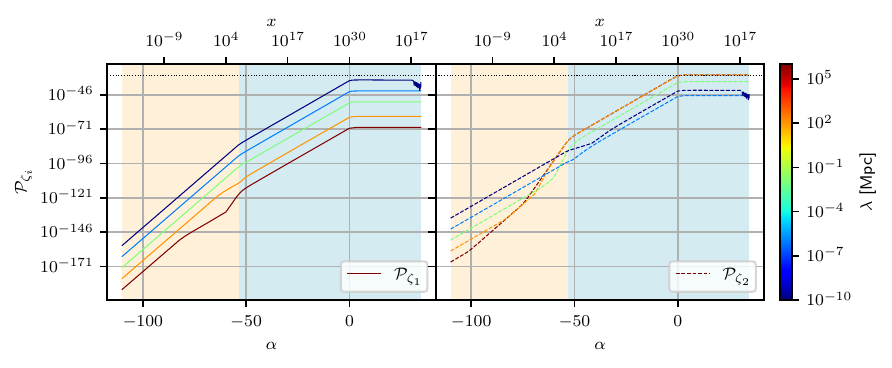}
  \caption{Time evolution of the curvature perturbation power spectrum
    $\mathcal{P}_\zeta(k)$, showing Mode~1 contributions (solid curves) and Mode~2
    contributions (dashed curves). The color bar indicates comoving wavelength
    $\rwavelength$ (in Mpc), with the top axis showing the corresponding scale factor
    $x = a_0/a$. Three key features emerge from the evolution: First, all modes begin
    with a characteristic $k^2$ (blue-tilted) power spectrum, apparent from the larger
    amplitudes at smaller wavelengths. Second, Mode~2 displays irregular evolution -
    longer wavelengths that exit the SHS during matter domination (brown/red curves)
    undergo significant amplitude enhancement compared to modes that exited during
    radiation domination. Third, Mode~1 maintains its blue-tilted spectral shape
    throughout the entire evolution, showing no comparable amplification. These
    distinct behaviors reflect fundamentally different dynamics dependent on the SHS
    crossing time. The dotted line shows the approximate analytic result from
    Eq.~\eqref{Pk-zeta-approx} evaluated at $k=1/\RH$.}
  \label{fig:zeta_power_spectrum}
\end{figure*}

\subsection{Around the Bounce}

Near the bounce, using the time gauge defined in Eq.~\eqref{tau:gauge}, the Hubble
function behaves as $H \propto \tilde{\tau}$. From the definitions of $F_\nu$ and
$\gw_i$, which contain $H$ in the denominator, one then finds
\begin{equation}\label{gw:F:bounce}
  F_\nu \approx \frac{F_\nu^\textsc{b}}{|\tilde{\tau}|},
  \qquad
  \gw_i \approx \frac{\gw_i^\textsc{b}}{|\tilde{\tau}|},
\end{equation}
where $F_\nu^\textsc{b} \equiv \lim_{\tilde{\tau} \to 0} \qty(|\tilde{\tau}|F_\nu)$ and
$\gw_i^\textsc{b} \equiv \lim_{\tilde{\tau} \to 0} \qty(|\tilde{\tau}|\gw_i)$ are
finite constants. Substituting this scaling into Eqs.~\eqref{TFHamilton}, one sees that
$\tilde{\tau}=0$ is a regular singular point of the original system.

This apparent singularity originates from the definition of the canonical momentum
$\Pi_{\zeta k}$, Eq.~\eqref{Pizeta}, which introduces an explicit factor of
$1/\tilde{\tau}$. Substituting the rescaling $\Pi_{\zeta k} = P_{\zeta k} /
  \tilde{\tau} $ defined in Eq.~\eqref{Pzeta:rescaled}, the system can be rewritten in
terms of $(\zeta_k, Q_k, P_{\zeta k}, \Pi_{Qk})$. To leading order near the bounce, the
resulting equations reduce to a coupled second-order system with constant coefficients
(see Appendix~\ref{app:bounce}), for which $\tilde{\tau}=0$ is an ordinary point.
Consequently, all perturbation variables are regular at the bounce and admit Taylor
expansions around $\tilde{\tau}=0$; no divergence occurs in this class of solutions.
Finally, using Eq.~\eqref{def:Psi}, one finds that $\Psi_k \propto H\Pi_{\zeta k}$.
Since $H\propto\tilde{\tau}$ near the bounce and the rescaled canonical variables
remain finite, the combination $H\Pi_{\zeta k}$ is finite, and therefore $\Psi_k$ is
also regular at $\tilde{\tau}=0$.

\section{Results}

We now present the results of the numerical integration of the coupled system of
perturbation equations \eqref{TFHamilton}, using the initial conditions derived in
Appendix~\ref{app:adiabatic-approximation}. Our analysis follows the evolution of
perturbative modes through the contracting phase, across the bounce, and into the
expanding phase. We pay particular attention to the resulting power spectra and their
dependence on the model parameters.

\subsection{Evolution of Perturbative Modes}
\label{sec:mode_evol}

During the contracting phase, the background evolves from a dust-dominated regime to a
radiation-dominated one. At the perturbative level, this implies that comoving
wavelengths $\rwavelength = 1/k$ exit the WKB regime at different stages: some during
dust domination, others during radiation domination, and yet others during the
transition between the two. In Appendix~\ref{app:adiabatic-approximation}, we
established that the WKB scale is approximately given by $c_i F_\nu$.

Figure~\ref{fig:mode_evolution} illustrates this behavior for a range of modes. The top
panels compare each mode's comoving wavelength with the corresponding SHS, $c_i x R_H$,
making it easy to identify when modes leave the WKB regime and transition from sub- to
super-Hubble evolution. The bottom panels show the time evolution of $\zeta_1$ and
$\zeta_2$, highlighting how the oscillatory behavior is suppressed once modes exit the
SHS. Moreover, the WKB approximation plotted with dashed lines provides a reasonable
description of modes in the WKB regime. Modes that exit during dust
domination, radiation domination, or the transition exhibit the expected differences in
phase and amplitude, consistent with the relative values of $c_i F_\nu$ and the
gravitational weights $\gw_i$ discussed in Sec.~\ref{sec:mode_evol}.

Naturally, since we have two fluids, the curvature perturbation $\zeta$ does not
contain all the information about the system, and we also need to analyze the velocity
entropy perturbation $Q$. As defined in Eq.~\eqref{Smode}, this variable is
proportional to the difference between the curvature perturbations of each fluid:
\begin{equation}
  Q_i \propto \Delta\zeta_i \equiv \zeta_{\mathrm{r}i} - \zeta_{w i}.
\end{equation}

Figure~\ref{fig:mode_dzeta_evolution} shows the evolution of $\Delta\zeta_i$, analogous
to Fig.~\ref{fig:mode_evolution} for the curvature perturbations $\zeta_i$. The
qualitative behavior is similar: modes oscillate inside the SHS and evolve in a scale
independent way once they exit. However, the amplitudes and phases differ, reflecting
the distinct dynamics of isocurvature perturbations compared to curvature ones.
Comparing Figs.~\ref{fig:mode_evolution} and \ref{fig:mode_dzeta_evolution}, one sees
that both $\zeta_i$ and $\Delta\zeta_i$ grow during the contracting phase, leading to a
corresponding perturbation in the fluid velocities. This illustrates how the coupled
two-fluid system naturally generates both curvature and isocurvature perturbations
during contraction.

Now, the momenta $\Pi_{\zeta_i}$ and $\Pi_{Q_i}$ are also relevant, as they
characterize the dynamics of the perturbation energy densities. They are given by
Eqs.~\eqref{Pizeta} and \eqref{PiS}, respectively. Using these expressions, we compute
the total energy density contrast $\ec_{\rho}$, Eq.~\eqref{ec_rho}, and the difference
between the individual fluid contrasts, $\ec_{\rho\mathrm{r}} - \ec_{\rho w}$, which is
given directly by the momenta $\Pi_{Q_i}$. Figure~\ref{fig:mode_drho_evolution} shows
the evolution of $\ec_{\rho}$, which exhibits oscillations inside the SHS and freezes
outside, similarly to $\zeta_i$ and $\Delta\zeta_i$. The amplitude of the total energy
density grows during contraction, indicating that the energy density perturbations
become significant.

By contrast, the isocurvature perturbations, defined as the difference between the
fluid energy density contrasts, are dynamically suppressed relative to the curvature
ones, as shown in Fig.~\ref{fig:mode_ddelta_rho_evolution}. While they also oscillate
inside the SHS and evolve in a scale independent way, their amplitude remains much
smaller than that of the total energy density contrast. Thus, although entropy
perturbations are generated during contraction, they do not grow significantly. This
behavior is consistent with the expectation that the curvature perturbations dominate,
while isocurvature modes remain subdominant. A direct comparison between
Figs.~\ref{fig:mode_drho_evolution} and \ref{fig:mode_ddelta_rho_evolution} shows that
the isocurvature modes are suppressed by several orders of magnitude relative to the
curvature ones.

Having analyzed all perturbative modes, we can now summarize their dynamics. Mode~1
evolves according to the SHS $c_\mathrm{r} x \RH$, while Mode~2 follows $c_w x \RH$. In
the WKB regime, their amplitudes generally grow (with the exception of $\Delta\ec_{\rho
    1}$), as seen from the leading terms in Eqs.~\eqref{ev1} and \eqref{ev2}.

The canonical momentum $\Pi_{Q1}$, however, behaves differently. Its
leading WKB amplitude is
\begin{equation}
  \left|\bar{\Pi}{Q1}\right| = \sqrt{\frac{F\nu}{2c\rr\gw\rr}} =
  \sqrt{\frac{3k}{8c\rr\Omega\rr}},
\end{equation}
where we have used Eqs.~\eqref{Fnu} and~\eqref{gw} to substitute the gravitational
weight $\gw\rr$. Unlike other perturbations, $\bar{\Pi}_{Q1}$ remains constant
throughout the WKB regime, implying a time-independent contribution to the entropy
perturbation during this phase.

Thus, while all other perturbation modes grow during sub-Hubble evolution in the
contracting phase, $\Pi_{Q1}$ stays approximately constant. After SHS exit, all modes
enter the super-Hubble regime. An exception is $\Pi_{Q2}$, which exhibits a transient
evolution before reaching its asymptotic constant value. This illustrates the super-SHS
dynamics discussed earlier, where early-time effects and the action of the evolution
matrix $M$ can shift the constant part of the solution, particularly for modes
sensitive to initial conditions.

\begin{figure*}
  \centering
  \includegraphics[width=0.99\textwidth]{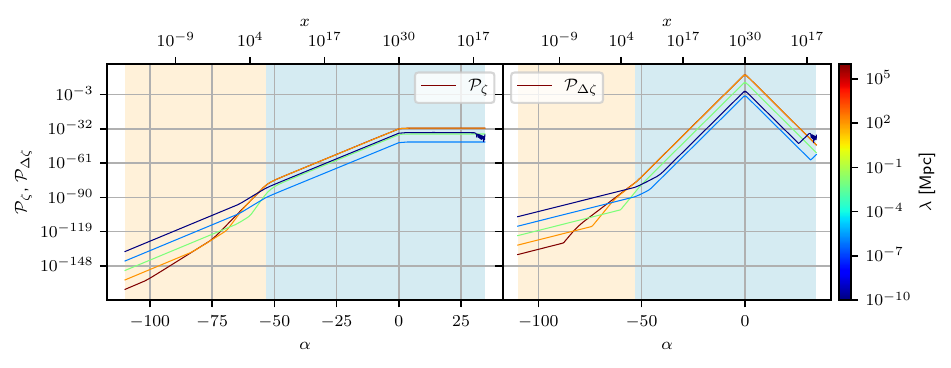}
  \caption{Time evolution of the total curvature perturbation power spectrum
    $\mathcal{P}_\zeta(k)$ (left panel) and the associated isocurvature spectrum
    $\mathcal{P}_{\Delta\zeta}(k)$ (right panel), combining contributions from both
    Mode~1 and Mode~2. The color bar indicates the comoving wavelength $\rwavelength$
    (in Mpc), and the top axis shows the corresponding scale factor $x = a_0/a$. The
    curvature spectrum $\mathcal{P}_\zeta(k)$ evolves from an initially blue-tilted
    $k^2$ shape to a final form characterized by a nearly scale-invariant plateau at
    large scales (long wavelengths) and a pronounced decline at small scales (short
    wavelengths). This evolution reflects the different amplification of modes
    depending on their SHS crossing epoch: modes exiting during matter domination are
    amplified to form the plateau, while those exiting during radiation domination
    retain their blue tilt with reduced amplitude. The resulting spectrum encodes the
    interplay between the two fluid components during the contracting phase. The
    isocurvature spectrum $\mathcal{P}_{\Delta\zeta}(k)$ also begins with a blue tilt
    but evolves into a mixed spectrum at late times. Although it can temporarily exceed
    the curvature spectrum around the bounce, its amplitude decreases afterward,
    eventually becoming smaller than $\mathcal{P}_\zeta(k)$.}
  \label{fig:zeta_total_power_spectrum}
\end{figure*}

\subsection{Power Spectra Evolution}

Our analysis of individual perturbation modes now leads us to their observable
consequences, namely the primordial power spectra that govern cosmic structure
formation. These spectra encode the statistical properties of three fundamental
quantities: the curvature perturbation $\zeta$ that seeds large-scale structure
formation, the velocity isocurvature perturbation $\Delta\zeta = \zeta_r-\zeta_w$ that
captures fluctuations between radiation and matter components, and the energy density
contrasts $\ec_{\rho}$ and $\ec_{\rho\mathrm{r}} - \ec_{\rho w}$ that describe energy
density isocurvature perturbations.

In Fig.~\ref{fig:zeta_power_spectrum}, we show the time evolution of the curvature
perturbation power spectrum $\mathcal{P}_\zeta(k)$, distinguishing between Mode~1
(solid curves) and Mode~2 (dashed curves). Three key features characterize the
dynamics: First, all modes initially follow a $k^2$ (blue-tilted) power spectrum,
evidenced by the enhanced amplitudes at smaller wavelengths. Second, Mode~2 exhibits
scale-dependent evolution: longer wavelengths crossing the SHS during matter domination
(brown/red curves) undergo significant amplification compared to those crossing during
radiation domination (blue/green curves). After SHS crossing, these modes asymptote to
a nearly scale-invariant spectrum.

In contrast, Mode~1 preserves its blue-tilted spectral shape throughout the evolution.
Since shorter wavelengths of Mode~2 remain unamplified, the total spectrum becomes
Mode~1-dominated at small scales. This contrast between modes, with their amplification
sensitive to the SHS crossing epoch, demonstrates the necessity of including both
contributions to fully capture the curvature perturbation's dynamics.

In order to help interpret these results, in Fig.~\ref{fig:zeta_total_power_spectrum}
we show the evolution of the total curvature perturbation power spectrum
$\mathcal{P}_\zeta(k)$. Note that large wavelengths leave the SHS during dust
domination and are amplified and join in an almost scale invariant spectrum, while
small wavelengths leave during radiation domination and remain blue tilted. However,
since the latter have a smaller amplitude, the final effect is a large decrease in
power for small scales followed by a blue spectrum.

The right panel of Fig.~\ref{fig:zeta_total_power_spectrum} shows the evolution of the
isocurvature perturbation power spectrum $\mathcal{P}_{\Delta\zeta}(k)$. Similar to the
curvature spectrum, it is initially blue-tilted and evolves into a mixed spectrum at
late times. The overall amplitude remains suppressed relative to the curvature power
spectrum, as expected. One might question whether the temporary enhancement of the
isocurvature amplitude around the bounce could indicate a breakdown of the linear
approximation. However, as discussed in~\cite{vitenti2012large}, linear theory remains
valid as long as perturbations are small in the chosen gauge. In
Sec.~\ref{sec:isocurvature}, we show that in the constant-curvature gauge the
isocurvature perturbation remains linear through the bounce and becomes further
suppressed afterward.

\subsection{Curvature Power Spectrum}
\label{sec:adiabatic-pz}

\begin{figure}
  \centering
  \includegraphics[scale=1.0]{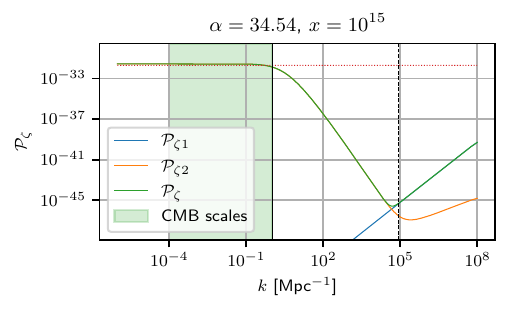}
  \caption{Power spectrum $\mathcal{P}_{\zeta}(k)$ of adiabatic perturbations in the
    two-fluid model. The spectrum exhibits two approximate power-law regimes associated
    with matter and radiation domination, separated by a transition interval
    $k^\textsc{s}_2 < k < k^\text{eq}_2$. These scales are shown as vertical lines in
    the plot: $k^\textsc{s}_2$ (solid) and $k^\mathrm{eq}_2$ (dashed). The CMB-relevant
    scales ($10^{-4}\,\mathrm{Mpc}^{-1} < k < 1\,\mathrm{Mpc}^{-1}$) are highlighted in
    green.}
  \label{fig:spec}
\end{figure}

We present the adiabatic curvature power spectrum $\mathcal{P}_{\zeta}(k)$ for the
two-fluid model, computed numerically and shown in Fig.~\ref{fig:spec}. At long
wavelengths, the analytic approximation in Eq.~\eqref{Pk-zeta-approx} provides a
reasonable estimate of both amplitude and spectral index,
\begin{equation}
  n_\textsc{s}\left( w \right) = 1 + \frac{12w}{1 + 3w},
\end{equation}
predicting an almost scale-invariant spectrum in this regime.

For shorter wavelengths, the SHS crossing occurs closer to the matter-radiation
transition, leading to a suppression of the amplitude before the spectral index reaches
$n_\textsc{s} = 3$ (see Eq.~\eqref{ps-mode-1}) and producing an overall red tilt in the
mean spectrum. This suppression begins at the scale $k^\textsc{s}_2$, Eq.~\eqref{defkS2}, when the sound speed in the matter-dominated era starts to
increase (from $\sqrt{w}$ to $1/\sqrt{3}$), and ends at the scale $k^\mathrm{eq}_2$,
Eq.~\eqref{defkeq2}, corresponding to the background transition from matter to
radiation domination. These two scales are indicated as vertical lines in
Fig.~\ref{fig:spec}, where one can see that both scales correctly mark the transition
for the power spectrum shape.

Since the curvature spectrum becomes constant shortly after the bounce, the precise
time at which it is evaluated is irrelevant. In contrast, as we will see in the next
section, the isocurvature power spectrum generally depends on time, requiring a
specific evaluation epoch. Here we adopt $x = 1 + z = 10^{15}$, corresponding to a time
well after the bounce and well above the nucleosynthesis redshift ($z \simeq 10^{10}$).

\begin{figure}[ht]
  \centering
  \includegraphics[scale=1.0]{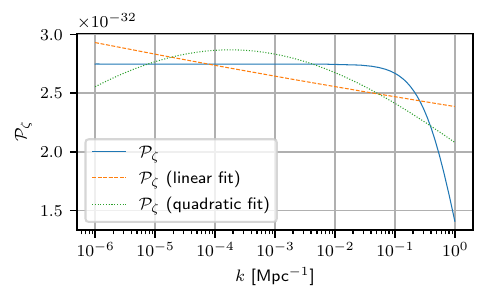}
  \caption{Figure showing the curvature power spectrum for $\mathcal{P}_{\zeta}(k)$,
    localized for the CMB $10^{-6} < k < 1$. We included a linear and quadratic fits,
    resulting in $n_\textsc{s} = 0.9851$ and $n_\textsc{s} = 0.9514$ with running
    parameter $\alpha_s = -0.0086$.}
  \label{fig:zeta:ns}
\end{figure}

In Fig.~\ref{fig:zeta:ns}, we focus on scales relevant for the CMB and quantify the
spectral shape using both linear and quadratic fits in $\ln k$. We adopt the pivot
scale $k_0 = 0.05\,\mathrm{Mpc}^{-1}$, consistent with
Planck~2018~\cite{planck2018_results}, and parametrize the spectrum as
\begin{equation}
  n_\textsc{s}(k) = n_\text{eff} + \frac{1}{2} \alpha_s
  \ln\frac{k}{k_0},
\end{equation}
where $\alpha_s \equiv \dd n_\text{s}/\dd\ln k$ is the running of the
spectral index.  The fits yield
\begin{itemize}
  \item Linear (constant $n_\textsc{s}$): $n_\text{eff} = 0.9851$.
  \item Quadratic (allowing for running): $n_\text{eff} = 0.9514$,
        $\alpha_s = -0.0086$.
\end{itemize}
The running captures the mild scale dependence across the CMB scales ($k \lesssim
  1\,\mathrm{Mpc}^{-1}$), mainly due to the loss of amplitude at short scales. While
these numbers are close to those obtained by Planck, they arise from fitting a
spectrum that is not inherently polynomial; thus, a quadratic approximation is only
indicative. A proper comparison with observations requires a full statistical
analysis using the numerical spectrum directly.

At first glance, the fact that the spectrum is not a simple power law with
$n_\textsc{s} \approx 0.96$ might seem like a shortcoming of the model. However, as
emphasized by the Planck collaboration \cite{planck2018_inflation}, a model-independent
reconstruction of the primordial power spectrum from CMB data does not require a
red-tilted power law.\footnote{A power law with $n_\textsc{s} \approx 0.96$ is
  \textit{consistent} with current observations, but not uniquely preferred.} Indeed,
examining Fig.~20 of \cite{planck2018_inflation}, one sees that a spectrum that is
approximately scale-invariant at large scales with a mild suppression of power at
smaller scales appears to be compatible with the data. A definitive assessment,
however, requires a full statistical analysis and, ideally, a model comparison using
the numerical spectrum.

\begin{figure}[ht]
  \centering
  \includegraphics[scale=0.98]{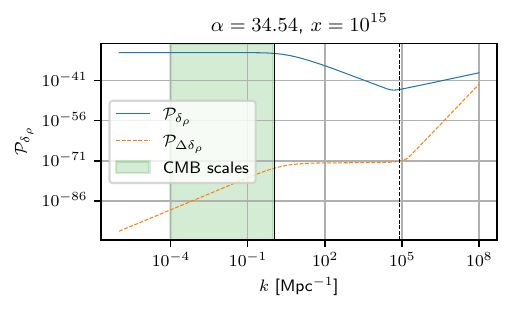}
  \caption{Power spectrum of isocurvature perturbations in the two-fluid model,
    computed at $x = 10^{15}$ during the expanding phase. The spectrum, obtained from
    Eqs.~\eqref{ec_rho} and~\eqref{PiS}, shows that the isocurvature amplitude is
    strongly suppressed relative to the curvature spectrum, although it approaches it
    at very small scales due to its steeper slope.}
  \label{fig:isocurvature}
\end{figure}

\subsection{Isocurvature Perturbations}
\label{sec:isocurvature}

To compute the CMB temperature power spectrum, we must also consider possible
isocurvature perturbations. In standard single-field inflation, such perturbations are
not generated during the inflationary phase, since the dynamics is governed by a
single degree of freedom, but can be produced after inflation, e.g. during the
reheating phase, when different components may acquire distinct density fluctuations.
In contrast, multi-field inflationary models naturally allow isocurvature modes to be
generated already during inflation. Current observations from
Planck~\cite{planck2018_results} show that isocurvature contributions to the CMB power
spectrum are very small: they have only been constrained, not measured.

In the two-fluid model, the isocurvature perturbations are obtained from the total
energy density~\eqref{ec_rho} and its associated perturbation~\eqref{PiS}. The
resulting power spectrum is shown in Fig.~\ref{fig:isocurvature}. The plot indicates
that, for the quantities defined by Eqs.~\eqref{ec_rho} and~\eqref{PiS}, the
isocurvature amplitude is strongly suppressed compared to the curvature one.
Nonetheless, the isocurvature spectrum is steeper and approaches the curvature spectrum
at very small scales. This behavior implies that the near absence of large-scale
isocurvature perturbations is a natural prediction of the model, consistent with
observations, without invoking additional mechanisms such as those required during
reheating.

\begin{figure}[ht]
  \centering
  \includegraphics[scale=0.98]{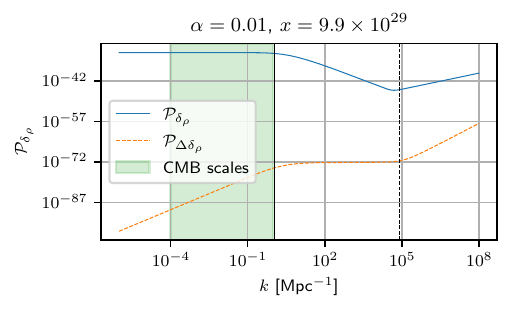}
  \caption{Power spectrum of isocurvature perturbations in the two-fluid model,
    computed near the bounce ($\alpha = 0.01$, $x \approx a_\textsc{b}$) during the
    expanding phase. As in Fig.~\ref{fig:isocurvature}, the isocurvature amplitude,
    derived from Eqs.~\eqref{ec_rho} and~\eqref{PiS}, remains strongly suppressed
    relative to the curvature spectrum, despite becoming comparable at very small
    scales due to its steeper slope. Even at the bounce, the isocurvature power
    spectrum remains well below the curvature one. As seen in
    Fig.~\ref{fig:isocurvature}, the main difference is that the isocurvature spectrum
    increases at smaller scales after the bounce, leading to a noticeably steeper
    dependence on $k$.}
  \label{fig:isocurvature_bounce}
\end{figure}

Note, however, that this isocurvature power spectrum is computed at the same time as
the adiabatic one ($x = 10^{15}$). To assess its time dependence, we also evaluate both
spectra at the bounce ($\alpha = 0.01$, $x \approx a_\textsc{b}$). The corresponding
power spectrum is shown in Fig.~\ref{fig:isocurvature_bounce}. As seen for $\ec_{\rho}$
and $\Delta\ec_{\rho}$, the isocurvature power spectrum remains smaller than the
curvature one at all times.  The behavior of the isocurvature power spectrum differs
significantly from that of the curvature perturbations. While the energy density
contrast $\ec_{\rho}$ remains suppressed at all times, the adiabatic perturbations
$\zeta$ are enhanced around the bounce, as shown in
Fig.~\ref{fig:zeta_total_power_spectrum}. In particular, the isocurvature power
spectrum can reach amplitudes $\Delta\zeta > 1$.

This can be understood by noting that $\zeta$ is a weighted combination of the
individual fluid perturbations $\zeta\ww$ and $\zeta\rr$, Eq.~\eqref{zeta}. Even if
$\zeta\ww$ grows very large, the total $\zeta$ can remain small during the radiation
dominated phase because $\gw\ww \ll \gw$, so that the combination $\gw\ww \zeta\ww /
  \gw$ is suppressed. As a result, $\zeta\ww$ grows rapidly during contraction and
decreases rapidly during expansion, while the total $\zeta$ remains moderate.

\begin{figure*}
  \centering
  \includegraphics[scale=0.98]{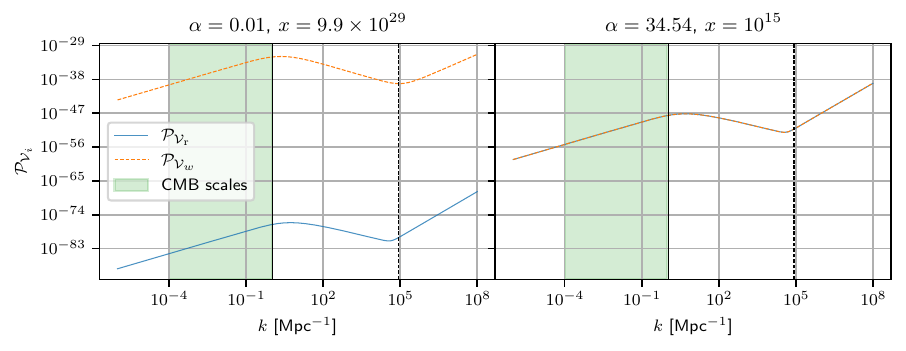}
  \caption{Power spectrum of the fluid velocity perturbations $kx\mathcal{V}\ww$ and
  $kx\mathcal{V}\rr$ in the two-fluid model (see Eq.~\eqref{Vpot}), computed near the
  bounce ($\alpha = 0.01$, $x \approx a_\textsc{b}$, left panel) and during the
  expanding phase ($x=10^{15}$, right panel). Near the bounce, $kx\mathcal{V}\ww$ is
  enhanced during contraction, while $kx\mathcal{V}\rr$ grows more slowly, producing a
  significant difference in the power spectra. During the expanding phase, both
  velocity perturbations converge to the same spectrum, reflecting the suppression of
  isocurvature velocity perturbations.}
  \label{fig:isocurvature_velocity}
\end{figure*}

Evaluating these quantities in a specific gauge further clarifies the behavior around
the bounce and confirms that the linear regime is maintained. Using the physical
velocity perturbation defined in Eq.~\eqref{Vpot}, Fig.~\ref{fig:isocurvature_velocity}
shows that the isocurvature velocity perturbation is enhanced during contraction but
becomes suppressed during expansion, eventually vanishing while never exceeding unity.
The left panel illustrates that the matter isocurvature velocity perturbation grows
during contraction but remains below one, while the right panel shows that both
velocity potentials converge to the same spectrum during expansion, indicating the
strong suppression of isocurvature velocity perturbations in this phase. This behavior
is also evident at the very end of the evolution of $\Delta\zeta$ in
Fig.~\ref{fig:zeta_total_power_spectrum}, where it decreases below the curvature
spectrum at late times.

Finally, although the perturbations undergo a complex evolution during contraction, the
bounce, and the subsequent expansion, the resulting spectrum shown in
Fig.~\ref{fig:zeta:ns} is compatible with negligible isocurvature perturbations at the
onset of CMB evolution, at least for the scales relevant to the CMB power spectrum, as
we will see in the following sections. Consequently, the computation of the CMB angular
power spectrum can be done in the same way as what is done for an inflationary
scenario, namely using only the primordial power spectrum, here provided from
Fig.~\ref{fig:zeta:ns}. Although the correlation between curvature and isocurvature
perturbations can be strong in terms of the correlation coefficient, the actual
covariance is negligible due to the suppressed isocurvature amplitude. As a result,
cross-correlation terms do not significantly contribute to the CMB power spectrum and
can be safely ignored.

\subsection{Tensor Perturbations}
\label{sec:tensor}

Having established the behavior of scalar perturbations, including the red tilt and
power suppression at small scales, we now examine whether primordial gravitational
waves remain consistent with observational constraints. This requires analyzing tensor
perturbations and the tensor-to-scalar ratio $r$.

Tensor perturbations in two-fluid bouncing cosmologies were previously studied in
\cite{bessada2012stochastic} for radiation plus a perfect dust fluid with $w = 0$.
Since we consider an almost pressureless fluid with $w \ll 1$, we expect qualitatively
similar behavior, though the details differ due to our specific background evolution.

Tensor perturbations $h_{\lambda k}$ evolve according to the equations of motion
\cite{bessada2012stochastic, bacalhau2018consistent} (see also
Ref.~\cite{Micheli:2022tld})
\begin{equation}
  h^{\prime\prime}_{\lambda k} +
  2\frac{m_{h}^{\prime}}{m_{h}}h^{\prime}_{\lambda k} + \nu^2
  _{\textsc{t}k}h_{\lambda k} = 0\, ,
\end{equation}
where $\lambda = +, \times$ refers to the polarization of the tensor perturbation and
we introduced the effective masses and frequencies
\begin{subequations}
  \begin{align}
    m^2 _{h} & \equiv \frac{a^{3}}{N} \, , \\ \nu^2 _{\textsc{t}k} &
    \equiv \dpar{ \frac{Nk}{a} }^2 \, ,
  \end{align}
\end{subequations}
where $N$ is the lapse function~\cite{bacalhau2018consistent}.

\begin{figure}[!ht]
  \centering
  \includegraphics[width=0.48\textwidth]{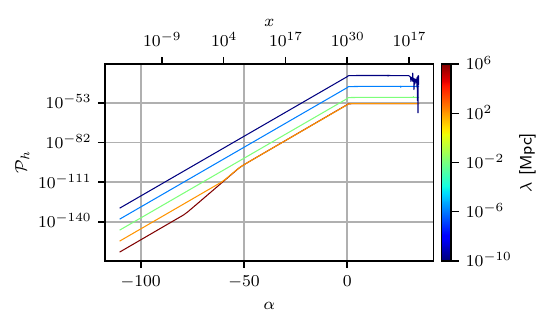}
  \caption{Time evolution of the tensor power spectrum $\mathcal{P}_h(k)$ for different
    comoving wavelengths $\rwavelength$ (in Mpc), with the corresponding scale factor
    $x = a_0/a$ shown on the top axis. Modes crossing the Hubble radius during matter
    domination (brown/red) exhibit different evolution from those exiting during
    radiation domination (blue/green), though both freeze out after horizon crossing.}
  \label{fig:tensor_pk_evol}
\end{figure}

Unlike scalar perturbations, where the two fluids contribute coupled modes, tensor
perturbations are decoupled and evolve independently for each polarization $\lambda =
  +, \times$. This decoupling allows us to quantize the system using a standard adiabatic
vacuum prescription, setting initial conditions deep in the contracting phase where
modes are well inside the Hubble radius. We then evolve the modes numerically through
the bounce and into the expanding phase, following the procedure detailed in
Ref.~\cite{bessada2012stochastic}.

The dimensionless tensor power spectrum is defined as
\begin{equation}
  \mathcal{P}_h (k) = \frac{1}{2\pi^2
  }k^{3}\sum_{\lambda = +, \times}|h_{\lambda k}|^2 \, ,
\end{equation}
where the sum accounts for both polarization states. Figure~\ref{fig:tensor_pk_evol}
shows the evolution of $\mathcal{P}_h(k)$ for representative modes. As in the scalar
case, the spectrum exhibits two distinct behaviors depending on whether modes cross the
Hubble radius during matter or radiation domination. However, here we have a crucial
difference, tensor perturbations propagate at the speed of light, whereas scalar
perturbations have a varying sound speed. Consequently, the transition between the two
regimes is governed solely by Hubble crossing rather than SHS crossing, leading to
qualitatively different dynamics.

\begin{figure}[!ht]
  \centering
  \includegraphics[width=0.48\textwidth]{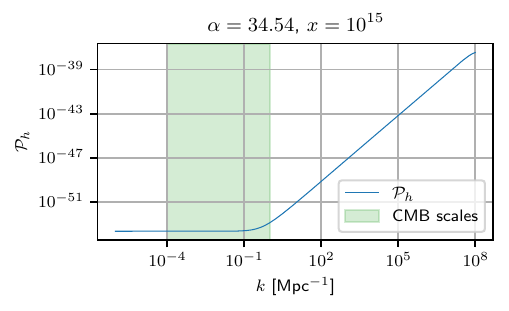}
  \caption{Tensor power spectrum $\mathcal{P}_h(k)$ computed at $x = 10^{15}$ during
    the expanding phase. The spectrum is nearly scale-invariant at large scales (CMB
    regime) and transitions to a steeper slope ($n_\textsc{t} \approx 3$) at smaller
    scales, with no suppression at the transition unlike the scalar spectrum.}
  \label{fig:tensor_pk}
\end{figure}

The final tensor power spectrum, computed at $x=10^{15}$ during the expanding phase and
shown in Fig.~\ref{fig:tensor_pk}, exhibits a nearly scale-invariant spectral index at
CMB scales, transitioning to a steeper slope ($n_\textsc{t} \approx 3$) at smaller
scales. Notably, unlike the scalar spectrum, there is no power suppression for modes
crossing during radiation domination, and the spectrum simply changes slope at the
transition scale.

\begin{figure}[!ht]
  \centering
  \includegraphics[width=0.48\textwidth]{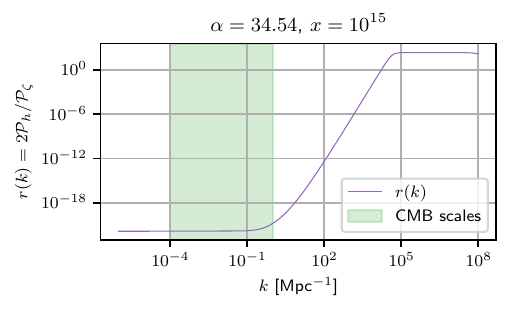}
  \caption{Tensor-to-scalar ratio $r(k)$ as a function of comoving wavelength
    $\rwavelength$ (in Mpc). At CMB scales, $r \sim 10^{-22}$, safely below Planck
    constraints \cite{planck2018_inflation}. The ratio increases dramatically at
    smaller scales, reaching $r(k) \sim 10^{3}$, where gravitational waves may be
    detectable by LISA or pulsar timing arrays.}
  \label{fig:tensor_ratio}
\end{figure}

To assess observational viability, we compute the tensor-to-scalar ratio,
defined as
\begin{equation}
  r(k) \equiv \frac{ \mathcal{P}_h(k) }{ \mathcal{P}_\zeta(k)
  },
\end{equation}
and plotted in Fig.~\ref{fig:tensor_ratio}. At CMB scales, the tensor amplitude is
substantially smaller than the scalar one, yielding $r \sim 10^{-22}$, which lies about
$22$ orders of magnitude beneath the Planck bound $r \lesssim 0.10$
\cite{planck2018_inflation}. This demonstrates that primordial gravitational waves do
not threaten the viability of the two-fluid model. Consequently, tensor perturbations
can be safely neglected when computing CMB temperature and polarization predictions.

While tensor perturbations are negligible at CMB scales, the situation changes at
smaller scales. As Figs.~\ref{fig:tensor_pk_evol} and~\ref{fig:tensor_ratio} show, the
tensor-to-scalar ratio increases significantly in this regime, reaching $r(k) \sim
  10^{3}$ for the smallest scales. This behavior arises from two factors: the continuous
growth of the tensor spectrum and the suppression of scalar power at small scales
discussed in Sec.~\ref{sec:isocurvature}. Beyond the transition scale, both spectra
have similar slopes, so $r(k)$ plateaus at this elevated value.

This enhancement suggests that the model's primordial gravitational waves could be
detectable by gravitational wave observatories sensitive to smaller wavelengths, such
as LISA or pulsar timing arrays like NanoGrav. We defer a detailed analysis of these
predictions to future work.

\subsection{Angular Power Spectra}

Having computed the primordial power spectra, we now outline how these results can be
connected to observations. Given the theoretical scope and length of the present work,
we do not attempt a full statistical analysis here. Instead, the goal of this section
is to demonstrate how the primordial curvature power spectrum produced by the model can
be efficiently parametrized and incorporated into standard cosmological pipelines.

Using this parametrization, we illustrate how the model can be interfaced with
Boltzmann codes and observational pipelines. As an example, we show the resulting
angular power spectra and report a representative best-fit to Planck~2018 data. A
complete statistical analysis, including full MCMC sampling and updated constraints,
will be presented in a forthcoming companion paper.

Our parametrization strategy exploits the analytical insights from
Sec.~\ref{sec:adiabatic-pz}. Recall that at large scales, the approximate expression in
Eq.~\eqref{Pk-zeta-approx} gives the amplitude at leading order in $w$ as
\begin{equation}\label{Pk_leading}
  \mathcal{P}_{\zeta}(k) \propto
  \frac{\lP^2x_\textsc{b}^2\Omega\ww^2}{\RH^2\Omega\rr c\ww^{5}}
  \equiv q_A,
\end{equation}
while the transition scale $k^\textsc{s}_2$ marking the onset of the red tilt
(see Eq.~\eqref{defkS2}) scales as
\begin{equation}\label{def:qk}
  \RH k^\textsc{s}_2 \propto \frac{\Omega\ww}{\sqrt{\Omega\rr}} \equiv q_k.
\end{equation}
These two quantities, illustrated in Fig.~\ref{fig:spec}, control the leading-order
dependence on model parameters: $q_A$ sets the overall amplitude, while $q_k$
determines the characteristic scale of spectral features. However, the full spectrum
also depends non-trivially on $w$ through the large-scale tilt and the detailed shape
of the transition region.

To capture this additional $w$-dependence efficiently, we precompute the full numerical
power spectrum on a dimensionless grid. Introducing the variable
\begin{equation}
  \kappa \equiv \RH^\textsc{f} k ,
\end{equation}
we evaluate the numerical spectrum
$\bar{\mathcal{P}}_{\zeta}(\kappa;w)$ for
$$
  10^{-9} \le w \le 10^{-1},
  \qquad
  10^{-3} \le \kappa \le 10^{8},
$$
using logarithmically spaced sampling in both variables. All remaining parameters are
fixed at fiducial values $\Omega\rr^\textsc{f}=10^{-5}$, $x_\textsc{b}^\textsc{f} =
  10^{30}$, and $H_0^\textsc{f}=70\,\mathrm{km}\,\mathrm{s}^{-1}\,\mathrm{Mpc}^{-1}$.

\begin{figure}[t]
  \centering
  \includegraphics[width=0.48\textwidth]{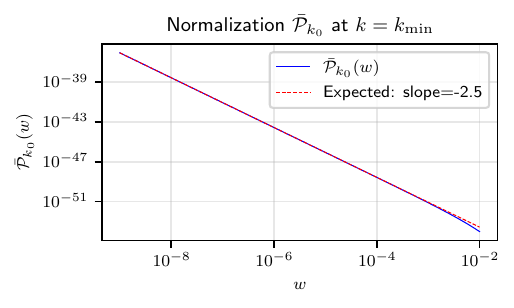}
  \caption{Scaling of $\bar{\mathcal{P}}_{\zeta}(\kappa=10^{-3};w)$ with $w$,
    verifying the expected $w^{-5/2}$ dependence. Deviations for
    $w \gtrsim 10^{-3}$ arise from higher-order corrections in $w$.}
  \label{fig:pk0_scaling}
\end{figure}

This scan serves two purposes. First, it allows us to verify the expected analytical
scaling. In particular, the spectrum evaluated at the fixed super-Hubble scale
$\kappa=10^{-3}$ follows the predicted power law
\begin{equation}
  \bar{\mathcal{P}}_{\zeta}(10^{-3};w) \propto w^{-5/2},
\end{equation}
as shown in Fig.~\ref{fig:pk0_scaling}. This confirms the leading-order behavior of the
amplitude factor $q_A$.

Second, this computation isolates the residual shape dependence of the spectrum. We
define the dimensionless shape function
\begin{equation}
  f(\kappa,w) \equiv
  \frac{\bar{\mathcal{P}}_{\zeta}(\kappa;w)}
  {\bar{\mathcal{P}}_{\zeta}(10^{-3};w)},
\end{equation}
which satisfies $f(10^{-3},w)=1$ by construction and depends only on the spectral
shape. In practice, $f(\kappa,w)$ is represented by a two-dimensional spline
interpolant in $(\log\kappa,\log w)$.

The primordial power spectrum can then be written as
\begin{equation}\label{Pk_parametrized_final}
  \mathcal{P}_{\zeta}(k; w, A_s, k_0)
  =
  A_s\, f\!\left(\frac{k}{k_0}, w\right),
\end{equation}
where the effective amplitude and scale parameters are
\begin{align}\label{As}
  A_s & =
  \bar{\mathcal{P}}_{\zeta}(10^{-3};w)\,\frac{q_A}{q_A^\textsc{f}}, \\
  \label{k0}
  k_0 & =
  \frac{q_k}{\RH\,q_k^\textsc{f}} .
\end{align}
This parametrization captures the dependence of the spectrum on $\Omega\rr$,
$x_\textsc{b}$ and $H_0$ through $(A_s,k_0)$, while the parameter $w$ controls the
remaining shape dependence encoded in $f$. We have verified that it reproduces the full
numerical spectrum to excellent accuracy throughout the parameter ranges of interest,
provided $w\ll1$ and the transition scale $k^\textsc{s}_2$ remains within the computed
domain.

Another relevant feature of this model is that, within the CMB window, the primordial
power spectrum probes only the onset of the transition between its two asymptotic
regimes. As shown in Figs.~\ref{fig:spec} and~\ref{fig:zeta:ns}, CMB scales span only
the initial part of the transition from the nearly scale-invariant large-scale regime
to the strongly blue-tilted small-scale regime. As a consequence, current CMB data are
not expected to place strong constraints on values of $w \ll 1$. Instead, they
primarily provide an upper bound on $w$ consistent with the observed spectra.

In this regime, the dominant sensitivity to $w$ enters through the amplitude parameter
$q_A$. However, this dependence is degenerate with that of other model parameters. By
contrast, the scale parameter $k_0$ is mainly sensitive to $\Omega\rr$, since $H_0$ is
already tightly constrained by CMB data alone. Therefore, while $k_0$ effectively
constrains $\Omega\rr$, the amplitude parameter $A_s$ constrains a combination of $w$
and $x_\textsc{b}$. This leads to a degeneracy in the $(w,x_\textsc{b})$ plane.

\begin{figure}[t]
  \centering
  \includegraphics[width=0.48\textwidth]{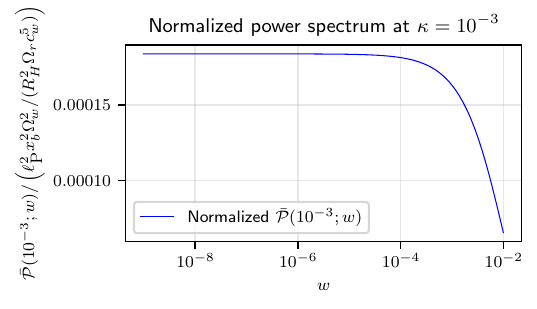}
  \caption{The normalized primordial power spectrum evaluated at $\kappa=10^{-3}$.
  Note that all parameters are fixed at their fiducial values except for $w$. For
  $w \lesssim 10^{-3}$, the spectrum follows the expected $w^{-5/2}$ scaling and
  this normalization factor $\mathcal{A}_0$ remains approximately constant, simplifying the
  inversion for $w$ from $A_s$.}
  \label{fig:pk0_normalized}
\end{figure}

For this reason, when translating the representative best-fit results into constraints
on the underlying physical parameters, we compute the translation as function of
$x_\textsc{b}$. The radiation density parameter $\Omega\rr$ can then be inferred from
the fitted scale parameter $k_0$ by inverting Eq.~\eqref{k0}, yielding
\begin{equation}
  \Omega\rr = \left(\frac{\sqrt{1+4B^2} - 1}{2 B}\right)^2,
  \qquad B \equiv \frac{1}{q_k^\textsc{f} k_0 \RH}.
\end{equation}
In order to invert for $w$, we need to use Eq.~\eqref{As} to express $w$ in terms of
$A_s$. However, there is a complicated dependence on $w$ through the shape function
$f$. Nonetheless, since $w$ is expected to be small in the CMB window, the
Fig.~\ref{fig:pk0_scaling} shows that for $w \lesssim 10^{-3}$, the factor $\mathcal{A}_0 \equiv
  \bar{\mathcal{P}}_{\zeta}(10^{-3};w)/q_A^\textsc{f}$ is approximately constant. For our
fiducial parameters, we find $\mathcal{A}_0 \approx 1.8 \times 10^{-4}$, one can see in
Fig~\ref{fig:pk0_normalized}. Given that, we can then write
\begin{equation}
  w \approx \left[ \frac{\mathcal{A}_0\left(x_\textsc{b}q_k^\textsc{f}\lP k_0\right)^2}{A_s} \right]^{2/5} ,
\end{equation}

In practice, the observational comparison is performed by fitting a standard
$\Lambda$CDM cosmological model, with the sole modification that the primordial
curvature power spectrum is replaced by the two-fluid spectrum described in this work.
All late-time physics—including the background expansion, recombination, and radiative
transfer—is treated as in $\Lambda$CDM, while the effects of the contracting two-fluid
phase enter exclusively through the primordial initial conditions encoded in
$\mathcal{P}_{\zeta}(k)$. This allows for a direct and consistent comparison with CMB
data using standard Boltzmann solvers.

Using the above parametrization, we obtain a representative fit to the Planck~2018 data
(baseline temperature and polarization: TT--TE--EE +
low-$\ell$)~\cite{planck2018_results}, yielding
\begin{subequations}\label{best_fit}
  \begin{align}
    H_0 & = 69.56\, \mathrm{km}\,\mathrm{s}^{-1}\,\mathrm{Mpc}^{-1}, \\
    A_s & = 2.16 \times 10^{-9},                                     \\
    k_0 & = 1.93 \times 10^{-3}\, \mathrm{Mpc}^{-1}.
  \end{align}
\end{subequations}
These values correspond to the inferred physical parameters
\begin{subequations}
  \begin{align}
    \Omega\rr & = 1.45 \times 10^{-7}, \\
    w         & = 9.44 \times 10^{-22}
    \left(\frac{x_\textsc{b}}{10^{30}}\right)^{4/5}.
  \end{align}
\end{subequations}
This, in turn, yields an asymmetry parameter
$\gamma_\textsf{assym} = 0.13$, consistent with the estimate obtained earlier.

\begin{figure}[t]
  \centering
  \includegraphics[width=0.48\textwidth]{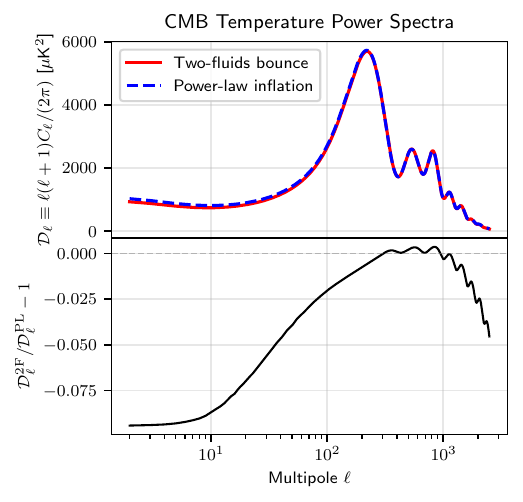}
  \caption{Comparison of the CMB angular power spectrum $\ell(\ell+1)C_{\ell}/(2\pi)$
  obtained from the two-fluid primordial spectrum (solid line) with that derived from a
  standard power-law spectrum (dashed line). Over most of the multipole range, the two
  models show good agreement, with the two-fluid model predicting slightly reduced
  power at the largest and smallest scales, at the level of at most $\sim 8\%$.}
  \label{fig:cmb_cls_comparison}
\end{figure}

In Fig.~\ref{fig:cmb_cls_comparison}, we show the resulting CMB angular power spectrum
$C_{\ell}$ computed using the CLASS Boltzmann solver~\cite{class} with the above
best-fit parameters, compared to the standard power-law primordial spectrum fitted to
the same data set.

Overall, this comparison demonstrates that the primordial power spectrum generated by
the contracting two-fluid model can be consistently embedded within the standard
$\Lambda$CDM framework and yields CMB angular power spectra compatible with current
observations. While a full statistical exploration of the parameter space is beyond the
scope of the present work, these results establish the phenomenological viability of
the model and provide a solid foundation for a dedicated follow-up analysis focusing on
parameter inference and observational constraints.

\section*{Conclusions}

In this work, we proposed and discussed an effective two fluid quantum cosmological
model in the framework of Canonical Quantum Gravity using a proper definition of
quantum trajectories, from which we extracted wavefunctions and their corresponding
non-singular quantum scale factor evolutions and perturbations. The two fluids we
considered are an almost (but not quite) pressureless fluid and radiation. We performed
an analysis for both background and perturbations, generalizing the literature on
bouncing models. At perturbative level, perturbations were quantized using the coupled
adiabatic vacuum prescription: Hamiltonian diagonalization followed by the adiabatic
expansion up to second order.

Concerning the predictions, at background level, the obtained scale factor is
non-singular and presents a bounce. At perturbative level the obtained adiabatic
primordial power spectrum presents a red-tilt, which is non-trivial to implement in
single fluid bouncing models. Furthermore, the contracting phase acts as a dynamical
mechanism that amplifies adiabatic perturbations without generating large entropy
perturbations, which explains the absence of isocurvature perturbations on the CMB
without fine-tuning nor need to postulate special initial conditions, as is often
performed in multi-field inflationary models.

Additional theoretical advantages arise due to the fact that, since this model does not
contain an inflationary phase, there is no need to postulate a reheating phase where
physics is unknown, nor are there any trans-Planckian frequencies that impact the CMB.
Furthermore, due to the use of the de Broglie-Bohm interpretation of quantum mechanics to extract trajectories associated to wavefunctions,
there is no quantum-to-classical transition problem.

Regarding observational constraints for this model, since the adiabatic power spectrum
is not an exact power law, one must perform a full MCMC analysis of CMB data to derive
the best fit parameters and confidence regions. This was originally intended to be
performed for this work, but the MCMC analysis of Planck Temperature + lowE data lead
to an interesting, non-trivial result: the Hubble constant value $H_{0}$ determined by
the CMB was increased using this power spectrum, so that the model offers an
alleviation for the well known Hubble Tension problem.

Because of these considerations, the MCMC analysis based on cosmological survey data is
presented separately in the companion paper Two Fluid Quantum Bouncing Cosmology Part
II: Observational Constraints~\cite{2fluid_pt2}. A preliminary assessment suggests that
the model can match current CMB and LSS observations in specific regions of parameter
space, while showing indications of reducing the $H_{0}$ and $\sigma_{8}$ tensions. If
these results are confirmed by the full analysis in Part II, the model may represent a
non-singular scenario capable of producing the primordial power spectrum without
invoking exotic matter, while remaining compatible with existing cosmological
constraints.

It should also be of note that, provided the bounce occurs rapidly enough, the
perturbative predictions of bouncing models differ only in the relation of the
amplitude to the bounce scale, $a_{\textsc{b}}$. Therefore, two fluid models with
different bouncing mechanisms, e.g. , Loop Quantum Cosmology, would lead to similar
predictions, defining a class of models that could be considered competitive
alternatives to inflationary models. Finally, if one considers the specific bouncing
mechanism of canonical quantization through the Wheeler-De Witt equation, this model
then leads to constraints on effective models of canonical quantum gravity using
cosmological data.

While the model already displays several interesting features, it is
still simple in the sense that it contains only two fluids; dark energy has not been
included, for example. Future work can then be based on the implementation of dark
energy in the contracting phase, which would demand still more advanced theoretical techniques, in particular
with regards to the definition of an appropriate vacuum state \cite{maier2012bouncing,
  Penna-Lima:2022dmx}. Furthermore, because the model remains consistent with the
constraints considered so far, further observational tests are also of interest. In
particular, observations that could probe the radiation domination region of the
spectrum where it grows due to the blue tilt become relevant. In this regard, spectral
distortions, the production of primordial black holes, primordial gravitational waves
and further LSS constraints seem promising, which we intend to address in future work.

\begin{acknowledgments}
  This work is supported by Conselho Nacional de Desenvolvimento Científico e
  Tecnológico (CNPq) -- Brasil. NPN acknowledges the support of CNPq of Brazil under
  grant PQ-IB 310121/2021-3. LFD acknowledges the support of CAPES under the grant DS 88887.902808/2023-00.
\end{acknowledgments}

\appendix

\section{Adiabatic Approximation for the Two-Fluid System}
\label{app:adiabatic-approximation}

In the contracting phase of our model, the perturbations are described by two coupled
degrees of freedom, $(\zeta, Q)$, associated with the two fluids. Quantizing this
system requires a choice of basis functions for the mode expansion. A natural starting
point is to identify an {\sl instantaneous} set of oscillatory solutions, so that each
mode can be followed as the background evolves. The adiabatic approximation provides a
systematic way to construct such a basis when the background changes slowly compared to
the oscillation timescale.

Defining the components of the phase-space vector by setting
\begin{equation}
  y^a \doteq \qty(\zeta, Q, \Pi_\zeta, \Pi_Q),
\end{equation}
we write the equations of motion in matrix form
\begin{equation}\label{2fluidmatrix}
  \dot{y}^a = M^a{}_b(\alpha)\, y^b ,
\end{equation}
with the evolution matrix $M^a{}_b(\alpha)$ given by
\begin{equation}
  \label{evolmatrix}
  M^a{}_b \doteq
  \begin{pmatrix}
    0              & 0                                & \dfrac{c_\zeta^2}{\gw} & \dfrac{s \Delta c^2 \gw\rr
    \gw\ww}{\gw^2}                                                                                          \\ 0 & 0 & \dfrac{s \Delta c^2 \gw\rr
    \gw\ww}{\gw^2} & \dfrac{\gw\ww \gw\rr c_Q^2}{\gw}                                                       \\ -\gw F_\nu^2
                   & 0                                & 0                      & 0                          \\ 0 & -\dfrac{\gw F_\nu^2}{\gw\rr \gw\ww} & 0 & 0
  \end{pmatrix}.
\end{equation}

If the background evolves on a timescale much longer than a typical oscillation period,
we may try a leading-order Wentzel-Kramers-Brillouin (WKB)-type solution:
\begin{equation}
  \bar{y}_i^{a}(\alpha) \equiv e^{-\ci\int \nu_i(\tilde{\alpha})\,
      \dd\tilde{\alpha}} \; v_i^a(\alpha),
\end{equation}
where $i=1,2$ labels two complex linearly independent solutions, $v_i^a(\alpha)$
representing the corresponding eigenvectors of the evolution matrix \eqref{evolmatrix}
whose eigenvalues $\nu_i(\alpha)\in\mathbb{R}$, given by
\begin{equation}
  \nu_1 = c\rr F_\nu,
  \qquad
  \nu_2 = c\ww F_\nu.
  \label{frequencies}
\end{equation}
play the role of instantaneous frequencies.

The instantaneous eigenmodes are thus defined by
\begin{equation}
  M^a{}_b(\alpha)\, v_i^b(\alpha) = -\ci\, \nu_i(\alpha)\,
  v_i^a(\alpha),
\end{equation}
(no sum) with
\begin{equation}
  v_i^a(\alpha) \doteq \qty(\bar{\zeta}_i, \bar{Q}_i, \bar{\Pi}_{\zeta
    i}, \bar{\Pi}_{Q i}).
\end{equation}
Substituting the ansatz into~\eqref{2fluidmatrix} yields
\begin{equation}
  \dot{\bar{y}}_i^a
  = \qty(-\ci\, \nu_i\, v_i^a + \dot{v}_i^a)
  e^{-\ci\int \nu_i\,\dd\alpha}
  = M^a{}_b\, v_i^b \, e^{-\ci\int \nu_i\,\dd\alpha}.
\end{equation}
At leading order in the adiabatic approximation, the variation of $v_i^a$ is neglected,
i.e. $\dot{v}_i^a \approx 0$. This corresponds to treating each eigenvector as fixed
while its phase oscillates with frequency $\nu_i$.

In reality, $M^a{}_b$ depends on background quantities and is therefore time-dependent,
so $v_i^a(\alpha)$ changes slowly. In the WKB expansion, this time variation generates
mode mixing through the correction term $\dot{v}_i^a$. The adiabatic approximation is
valid when these corrections are small: assuming a characteristic time $T$ for the
evolution of $v_i^a$, i.e. $[\dot{v}_i^a] \simeq v_i^a/T$, our adiabatic approximation
is valid provided $T\gg \nu_i^{-1}$, which requires that the background evolve on
timescales much longer than the oscillation period of each mode.

As shown in Ref.~\cite{Peter:2015zaa}, the matrix $M^a{}_b$ has two pairs of
complex-conjugate eigenvalues. Each pair corresponds to one of the two physical
oscillation frequencies \eqref{frequencies} of the coupled system: one propagating with
the speed-of-sound of the radiation component, $c\rr$, and the other with the
speed-of-sound of the cold-matter-like component, $c\ww=\sqrt{w}$. A convenient choice
of normalized eigenvectors (up to an overall phase) is
\begin{align}\label{ev1}
  v_1^a & \doteq \qty(s\sqrt{\frac{c\rr\gw\rr}{2F_\nu\gw^2}},
  \sqrt{\frac{c\rr\gw\rr}{2F_\nu\gw^2}}\gw\ww, -\ci s\sqrt{\frac{\gw\rr
      F_\nu}{2c\rr}}, -\ci\sqrt{\frac{F_\nu}{2c\rr\gw\rr}}),
\end{align}
and
\begin{align}
  \label{ev2}
  v^a_2 & \doteq \qty(-s\sqrt{\frac{c\ww\gw\ww}{2F_\nu\gw^2}},
  \sqrt{\frac{c\ww\gw\ww}{2F_\nu\gw^2}}\gw\rr, \ci s\sqrt{\frac{\gw\ww
      F_\nu}{2c\ww}}, -\ci\sqrt{\frac{F_\nu}{2c\ww\gw\ww}}).
\end{align}
The other two eigenvectors are the complex conjugates of the above, with respective
eigenvalues $-\nu_1$ and $-\nu_2$.

To compute products between eigenvectors (or between any solutions of the system), we
introduce the symplectic matrix
\begin{equation}
  \Omega_{ab} \doteq
  \begin{pmatrix}
    0  & 0  & 1 & 0 \\
    0  & 0  & 0 & 1 \\
    -1 & 0  & 0 & 0 \\
    0  & -1 & 0 & 0
  \end{pmatrix},
\end{equation}
which we use to define the Wronskian between $v_i$ and $v_j$ through
\begin{equation}
  W(v_i, v_j) \equiv v_i^{a*}\, \Omega_{ab}\, v_j^b,
\end{equation}
and the eigenvectors are normalized by the statement
\begin{equation}
  \label{wronskian-ij}
  W(v_i, v_j) = -\ci\delta_{ij},
\end{equation}
a condition that fixes the normalization for the quantum creation and annihilation
operators and, in the classical theory, provides a complete orthonormal basis.
Moreover, since the eigenvectors are mutually orthogonal, we have
\begin{equation}
  \label{wronskian-star-ij}
  W(v_i^*, v_j) = 0,
\end{equation}
The mode functions obtained by evolving these normalized eigenvectors and including the
adiabatic corrections described below reproduce exactly the vacuum choice of
Ref.~\cite{Peter:2015zaa}. Thus, the set $\{ v_k^a, v_k^{a*} \}$ forms a complete
instantaneous basis for the four-dimensional phase space in terms of which one can
decompose any vector.

To account for the slow time dependence of the eigenvectors $v_i^a(\alpha)$, we extend
the leading-order WKB ansatz by including a first-order correction
\begin{equation}
  y_i^a(\alpha) =
  e^{-\ci\int \nu_i(\tilde{\alpha})\, \dd\tilde{\alpha}}
  \left[ v_i^a(\alpha) + u_i^a(\alpha) \right],
\end{equation}
where $u_i^a(\alpha)$ describes the deviation from the instantaneous-eigenvector
solution induced by the slow evolution of the background.

Substituting this ansatz into the system~\eqref{2fluidmatrix} and keeping only
first-order terms in the adiabatic expansion gives
\begin{equation}
  \left( M^a{}_b + \ci \nu_i\, \delta^a{}_b \right) u_i^b(\alpha)
  = \dot{v}_i^a(\alpha) + \dot{u}_i^a(\alpha)
  \;\approx\; \dot{v}_i^a(\alpha),
  \label{kernel}
\end{equation}
where, in the last step, we neglected $\dot{u}_i^a$ since it is of second order in the
adiabatic parameter. The time derivative of the instantaneous eigenvector,
$\dot{v}_i^a(\alpha)$, therefore acts as a known source term for the correction
$u_i^a(\alpha)$.


Obtaining the first order correction $u_i^a(\alpha)$ through Eq.~\eqref{kernel}
requires to invert the operator $M^a{}_b + i \nu_i \delta^a{}_b$. Seen as acting in the
full 4-dimensional space of configurations, this operator is however not invertible
since the instantaneous eigenvector $v_i^a$ lies in its kernel. The three remaining
eigenvectors $v_{k\not= i}^a$ and $v_k^{a*}$, i.e. the other positive-frequency mode
and the two negative-frequency modes, are also eigenvectors of this operator, with
respective eigenvalues $-\ci \qty( \nu_k - \nu_i)$ and $\ci \qty( \nu_k + \nu_i)$, and
are linearly independent. They span the three-dimensional subspace orthogonal to
$v_i^a$ with respect to the symplectic product, subspace in which the source term $\dot
  v_i$ should lie, so we also demand $W(\dot v_i, v_i) = 0$, a condition that can be
enforced by an appropriate choice of phase for $v_i^a$.

To express this orthogonality condition explicitly, we exploit the fact that the set
$\{ v_k^a, v_k^{a*} \}$ forms a complete instantaneous basis for the four-dimensional
phase space. We can thus decompose\footnote{Note we write the coefficients as
  $\alpha_{ij}$ and $\beta_{ij}$ to match with the usual notation for such a Bogoliubov
  expansion, but the first coefficient should not be confused with our time parameter
  $\alpha$.} the time derivative of $v_i^a$ as
\begin{equation}
  \dot{v}_i^a = \alpha_i{}^k\, v_k^a + \beta_i{}^k\, v_k^{a*},
\end{equation}
where the sum over $k$ runs over the two frequency modes.

Taking the derivative of the orthonormality conditions Eqs.~\eqref{wronskian-ij} and
\eqref{wronskian-star-ij} with respect to $\alpha$ and substituting the above
decomposition yields the constraints
\begin{equation}\label{alpha-beta}
  \alpha_{ij}^* + \alpha_{ji} = 0,
  \qquad
  \beta_{ij} - \beta_{ji} = 0.
\end{equation}
Since $\alpha_{ii}$ is purely imaginary, it represents a phase rotation that can be
absorbed by redefining the phase of $v_i^a$. Moreover, we impose the gauge choice
$\alpha_{ii} = 0$, ensuring $W(\dot v_i, v_i) = 0$. The remaining coefficients
$\alpha_{ij}$ ($i \neq j$) and $\beta_{ij}$ describe physical mode mixing and particle
production.

When the first-order correction $u_i^a$ is included, the orthonormality conditions
\eqref{wronskian-ij} and \eqref{wronskian-star-ij} applied to $y_i^a = v_i^a + u_i^a$
imply, to first order,
\begin{align}
  W(u_i, v_j) + W(v_i, u_j) \approx 0
\end{align}
and
\begin{align}
  W(u_i^*, v_j) + W(v_i^*, u_j)\approx 0.
\end{align}
Decomposing $u_i^a=\mathfrak{a}_i^k v_k^a+\mathfrak{b}_i^k v_k^{*a}$, one recovers
Eq.~\eqref{alpha-beta} with the substitution $\alpha_{ij}\to \mathfrak{a}_{ij}$ and
$\beta_{ij}\to \mathfrak{b}_{ij}$. In particular, $\mathfrak{a}_{ii}$ is purely
imaginary and can be absorbed by a (first order) phase redefinition, leaving $W(u_i,
  v_i) = 0$. Thus both $\dot{v}_i^a$ and $u_i^a$ lie in the three-dimensional subspace
orthogonal to $v_i^a$, i.e. $W(\dot{v}_i, v_i) = 0$ and $W(u_i, v_i) = 0$.

Since both $\dot{v}_i^a$ and $u_i^a$ lie in the orthogonal subspace, we can explicitly
invert the operator on this subspace. \footnote{Naturally, if we apply the inverse to
the original operator, we obtain the projector onto the orthogonal subspace. That is,
the partition of the identity given by
$$
  \delta^a{}_b = -\ci \sum_{k=1}^2\left(v_k^{a*}v_k^c - v_k^av_k^{c*} \right)\Omega_{cb}.
$$
and the inverse operator satisfy
$$  \left(M^a{}_c + \ci \nu_i \delta^a{}_c\right)
  \left(M^c{}_b + \ci \nu_i \delta^c{}_b\right)^{-1}
  = \delta^a{}_b - \ci v_i^a v_i^{c*} \Omega_{cb}.
$$} For $i = 1$, the inverse takes the form
\begin{align}
  \qty(M^a{}_b + \ci \nu_1 \delta^a{}_b)^{-1} =
  -\qty( \frac{v^{a*}_1 v^c_{1}}{2 \nu_1}
  + \frac{v_2^av_2^{c*}}{\nu_2 - \nu_1}
  + \frac{v^{a*}_2 v^c_2}{\nu_2 + \nu_1})\Omega_{cb},
\end{align}
which is used to obtain $u^a_1(\alpha)$. This generalizes the standard WKB prefactor
for a single mode $1/(2\nu_i)$ and allows to include the full spectral structure, with
denominators given by differences between $\nu_1$ and the remaining eigenvalues.

\subsection{Explicit form of the first-order corrections}

To make the size of the first-order correction explicit, we factor it relative to each
component of the leading-order mode. For the $i$-th mode we write
\begin{align*}
  v_i^1 + u_i^1                   & = \bar{\zeta}_i \left( 1 + \delta_{\zeta_i} \right),
                                  & v_i^2 + u_i^2                                        & = \bar{Q}_i \left( 1 + \delta_{Q_i} \right),
  \\ v_i^3 + u_i^3 & = \bar{\Pi}_{\zeta i} \left( 1 +
  \delta_{\Pi_{\zeta i}} \right), & v_i^4 + u_i^4                                        & = \bar{\Pi}_{Q i}
  \left( 1 + \delta_{\Pi_{Q i}} \right),
\end{align*}
where $\delta_{\zeta_i}$, $\delta_{Q_i}$, $\delta_{\Pi_{\zeta i}}$ and $\delta_{\Pi_{Q
      i}}$ are complex functions of $\alpha$ describing the relative correction to each
component.  By construction, one has $\delta_{\zeta_i} \equiv u_i^1/v_i^1$,
$\delta_{Q_i} \equiv u_i^2/v_i^2$, $\delta_{\Pi_{\zeta i}} \equiv u_i^3/v_i^3$ and
$\delta_{\Pi_{Q i}} \equiv u_i^4/v_i^4$. This notation allows us to track amplitude and
phase shifts for each physical quantity ($\zeta, Q, \Pi_\zeta, \Pi_Q$) separately, and
to identify which components are most affected by adiabaticity violations. These
corrections depend on time derivatives of the background functions:
\begin{equation}
  \frac{\dot{\gw}_i}{\gw_i} = -3s\,c_i^2 + \frac{\dot{N}}{N}, \qquad
  \frac{\dot{F_\nu}}{F_\nu} = \frac{\dot{N}}{N} - s.
\end{equation}

Using these relations, the corrections for the first eigenmode are:
\begin{subequations}
  \label{eq:delta-1}
  \begin{align}
    \delta_{\zeta_1} & = \frac{1}{2\ci c\rr
      F_\nu}\left[\frac{s}{2}\left(3 c\rr^2 - 1 - \frac{6 c\rr^2
        \gw\rr}{\gw}\right)
      +\frac{\dot{N}}{N}\left(2-\frac{\gw\rr}{\gw}\right)\right],
    \\ \delta_{Q_1} & = \frac{1}{2\ci c\rr
      F_\nu}\left[\frac{s}{2}\left(3 c\rr^2 - 1 - \frac{6 c\rr^2
        \gw\rr}{\gw}\right)
      -\frac{\dot{N}}{N}\frac{\gw\rr}{\gw}\right],
    \\ \delta_{\Pi_{\zeta 1}} & = \frac{1}{2\ci c\rr
      F_\nu}\left[\frac{s}{2}\left(9 c\rr^2 + 1 - \frac{6 c\rr^2
        \gw\rr}{\gw}\right)
    -\frac{\dot{N}}{N}\frac{\gw\rr}{\gw}\right],                            \\ \delta_{\Pi_{Q
    1}}              & = \frac{1}{2\ci c\rr F_\nu}\left[\frac{s}{2}\left(-3
      c\rr^2 + 1 - \frac{6 c\rr^2 \gw\rr}{\gw}\right)
      -\frac{\dot{N}}{N}\frac{\gw\rr}{\gw}\right].
  \end{align}
\end{subequations}
and for the second eigenmode:
\begin{subequations}
  \label{eq:delta-2}
  \begin{align}
    \delta_{\zeta_2} & = \frac{1}{2\ci c\ww
      F_\nu}\left[\frac{s}{2}\left(-3 c\ww^2 - 1 + \frac{6 c\ww^2
        \gw\rr}{\gw}\right)
      +\frac{\dot{N}}{N}\left(1+\frac{\gw\rr}{\gw}\right)\right],
    \\ \delta_{Q_2} & = \frac{1}{2\ci c\ww
      F_\nu}\left[\frac{s}{2}\left(-3 c\ww^2 - 1 + \frac{6 c\ww^2
        \gw\rr}{\gw}\right)
      -\frac{\dot{N}}{N}\frac{\gw\rr}{\gw}\right],
    \\ \delta_{\Pi_{\zeta 2}} & = \frac{1}{2\ci c\ww
      F_\nu}\left[\frac{s}{2}\left(3 c\ww^2 + 1 + \frac{6 c\ww^2
        \gw\rr}{\gw}\right)
    -\frac{\dot{N}}{N}\frac{\gw\rr}{\gw}\right],                            \\ \delta_{\Pi_{Q
    2}}              & = \frac{1}{2\ci c\ww F_\nu}\left[\frac{s}{2}\left(-9
      c\ww^2 + 1 + \frac{6 c\ww^2 \gw\rr}{\gw}\right)
      -\frac{\dot{N}}{N}\frac{\gw\rr}{\gw}\right].
  \end{align}
\end{subequations}
The dominant parameter controlling the magnitude of the corrections above is $\qty(c_i
  F_\nu)^{-1}$, which can be rewritten as
$$
  \frac{1}{c_i F_\nu} = \frac{a|H|}{c_i k_c} =
  \frac{a\,\rwavelength_c}{c_i R_H}.
$$
Here, $\rwavelength \equiv 1/k$ is the comoving wavelength and $R_H \equiv 1/|H|$ is
the Hubble radius. The factor $c_i$ accounts for the propagation speed of the mode,
so this ratio measures the physical wavelength (including the sound speed) relative
to the Hubble radius.

This quantity determines the regime of validity of the WKB approximation: when $c_i
  F_\nu \gg 1$, the mode oscillates on a timescale much shorter than the Hubble time, and
the first-order corrections remain small.

Additionally, the remaining factors in the corrections are either of order unity or
involve the ratio $\gw\rr / \gw$. During the radiation-dominated phase this ratio is of
order unity, while during matter domination it is subdominant by definition: although
it grows in the contracting phase as $a^{-1+3w}\sim 1/a$, its amplitude remains small,
ensuring that the first-order corrections remain controlled and adiabaticity is
preserved at late times. After the transition to the radiation era, these factors
either stabilize at order unity or decay.

To estimate the truncation error of the adiabatic approximation, we extend the
perturbative expansion to include the second-order correction $q_i^a$,
$$
  y_i^a = v_i^a + u_i^a + q_i^a ,
$$
where $u_i^a$ is the first-order term and $q_i^a$ is the second-order term in the
adiabatic expansion. For example, for the first component,
$$
  v_i^1 + u_i^1 + q_i^1 = \bar{\zeta}_i \left[ 1 + \delta_{\zeta_i} +
    \delta_{\zeta_i}^{(2)} \right],
$$
where $\delta_{\zeta_i} \equiv u_i^1 / v_i^1$ is the first-order relative correction
and $\delta_{\zeta_i}^{(2)} \equiv q_i^1 / v_i^1$ is the second-order term. The same
definition applies to all other components.

The explicit forms of $\delta^{(2)}$ are lengthy and omitted here, but are implemented
in the numerical code used to generate the plots. In practice, we compute
$\delta_{\zeta_i}$ and $\delta_{\zeta_i}^{(2)}$ analytically, and estimate the
third-order term as the product $\delta_{\zeta_i}\,\delta_{\zeta_i}^{(2)}$.  Since the
adiabatic approximation must hold for all components simultaneously, we define the
adiabatic scale
\begin{equation}
  \label{eq:adiabatic-scale}
  \Delta_{\zeta_i} = \left| \delta_{\zeta_i} \, \delta_{\zeta_i}^{(2)} \right|^{1/3},
\end{equation}
with no summation over repeated $i$. This scale varies as $c_i F_\nu$, as discussed
earlier. The time at which the adiabatic approximation introduces a truncation error of
order $\epsilon$ is determined numerically by solving
$$
  \Delta_{\zeta_i}^3 + \Delta_{Q_i}^3 + \Delta_{\Pi_{\zeta i}}^3 +
  \Delta_{\Pi_{Q i}}^3 = \epsilon.
$$
This criterion provides a conservative estimate of the truncation error, and we have
verified through numerical integrations that it consistently overestimates the actual
error. From that point onward, the evolution is computed by directly integrating the
full coupled system.

\section{Regularity around the bounce}
\label{app:bounce}

Near the bounce, it is convenient to use the time gauge defined in
Eq.~\eqref{tau:gauge}. In this gauge, $H \propto \tilde{\tau}$, and the relevant
background functions have the following behavior: $F_\nu \approx
  F_\nu^\textsc{b}/|\tilde{\tau}|$ and $\gw_i \approx \gw_i^\textsc{b}/|\tilde{\tau}|$
(Eqs.~\eqref{gw:F:bounce}), where $F_\nu^\textsc{b}$ and $\gw_i^\textsc{b}$ are finite
constants. The sound speeds remain finite at the bounce.

Substituting this scaling into Eqs.~\eqref{TFHamilton} yields
\begin{subequations}
  \label{TFHamilton:tau2}
  \begin{align}
    \dv{\zeta_k}{\tilde{\tau}}       & = \tilde{\tau}^2\frac{c_\zeta^2}{\gw^\textsc{b}} \Pi_{\zeta k} + \tilde{\tau}
    \Delta c^2  \frac{\gw\rr^\textsc{b} \gw\ww^\textsc{b}}{\qty(\gw^\textsc{b})^2} \Pi_{Q k},                        \\
    \dv{Q_k}{\tilde{\tau}}           & = \frac{\gw\ww^\textsc{b} \gw\rr^\textsc{b} c_Q^2}{\gw^\textsc{b}} \Pi_{Q
      k} + \tilde{\tau} \Delta c^2  \frac{\gw\rr^\textsc{b} \gw\ww^\textsc{b}}{\qty(\gw^\textsc{b})^2} \Pi_{\zeta
    k},                                                                                                              \\
    \dv{\Pi_{\zeta k}}{\tilde{\tau}} & = -\frac{\gw^\textsc{b} F_\nu^{\textsc{b}2}}{\tilde{\tau}^2} \zeta_k,         \\
    \dv{\Pi_{Q k}}{\tilde{\tau}}     & = -\frac{\gw^\textsc{b} F_\nu^{\textsc{b}2}}{\gw\rr^\textsc{b}
      \gw\ww^\textsc{b}} Q_k,
  \end{align}
\end{subequations}
where $\tilde{\tau}=0$ is a regular singular point of the original system.

The apparent singularity originates from the definition of the canonical momentum
$\Pi_{\zeta k}$. Introducing the rescaled momentum
\begin{equation}\label{Pzeta:rescaled}
  P_{\zeta k} \equiv \tilde{\tau} \Pi_{\zeta k},
\end{equation}
the system becomes
\begin{subequations}
  \label{TFHamilton:tau3}
  \begin{align}
    \dv{\zeta_k}{\tilde{\tau}}                 & = \tilde{\tau}\frac{c_\zeta^2}{\gw^\textsc{b}} P_{\zeta k} + \tilde{\tau}
    \Delta c^2  \frac{\gw\rr^\textsc{b} \gw\ww^\textsc{b}}{\qty(\gw^\textsc{b})^2} \Pi_{Q k},                              \\
    \dv{Q_k}{\tilde{\tau}}                     & = \frac{\gw\ww^\textsc{b} \gw\rr^\textsc{b} c_Q^2}{\gw^\textsc{b}} \Pi_{Q
      k} + \Delta c^2  \frac{\gw\rr^\textsc{b} \gw\ww^\textsc{b}}{\qty(\gw^\textsc{b})^2} P_{\zeta
    k},                                                                                                                    \\
    \tilde{\tau}\dv{P_{\zeta k}}{\tilde{\tau}} & = P_{\zeta k} -\gw^\textsc{b} \qty(F_\nu^\textsc{b})^2 \zeta_k,           \\
    \dv{\Pi_{Q k}}{\tilde{\tau}}               & = -\frac{\gw^\textsc{b} \qty(F_\nu^\textsc{b})^2}{\gw\rr^\textsc{b}
      \gw\ww^\textsc{b}} Q_k.
  \end{align}
\end{subequations}

Differentiating the equations for the momenta, we find
\begin{subequations}
  \label{TFHamilton:tau4}
  \begin{align}
    \dv[2]{P_{\zeta k}}{\tilde{\tau}} & + \qty(F_\nu^\textsc{b})^2 \qty(c_\zeta^2 P_{\zeta k} +
    \Delta c^2  \frac{\gw\rr^\textsc{b} \gw\ww^\textsc{b}}{\gw^\textsc{b}} \Pi_{Q k})  = 0,     \\
    \dv[2]{Q_k}{\tilde{\tau}}         & + \qty(F_\nu^\textsc{b})^2 \qty(c_Q^2 \Pi_{Q
      k} +   \frac{\Delta c^2}{\gw^\textsc{b}} P_{\zeta
      k}) = 0.
  \end{align}
\end{subequations}
The system above has constant coefficients; therefore, $\tilde{\tau}=0$ is an ordinary
point of the reduced equations. Since the coefficients are constant, the solutions are
analytic and admit Taylor expansions around $\tilde{\tau}=0$. Substituting back into
Eqs.~\eqref{TFHamilton:tau3} shows that $\zeta_k$, $Q_k$, $\ec_\rho$ and $\Pi_Q$ --~see
Eqs.\eqref{Pizeta} and \eqref{PiS}~--, are all finite and admit Taylor expansions at
the bounce.

\bibliography{References}

\end{document}